\documentclass[acmsmall]{acmart}
\usepackage{mathrsfs}
\usepackage{multirow}
\usepackage{xspace}
\usepackage{subfigure} 
\usepackage{hyperref}
\usepackage{amsmath}
\usepackage{amsfonts}
\usepackage{xcolor}
\usepackage{epsfig}
\usepackage{diagbox}
\usepackage[ruled,linesnumbered]{algorithm2e}
\AtBeginDocument{%
  \providecommand\BibTeX{{%
    \normalfont B\kern-0.5em{\scshape i\kern-0.25em b}\kern-0.8em\TeX}}}

\setcopyright{acmcopyright}
\copyrightyear{2018}
\acmYear{2018}
\acmDOI{XXXXXXX.XXXXXXX}

\acmJournal{JACM}
\acmVolume{37}
\acmNumber{4}
\acmArticle{111}
\acmMonth{8}

\acmPrice{15.00}
\acmISBN{978-1-4503-XXXX-X/18/06}

\begin{document}

\title{DEWP: Deep Expansion Learning for Wind Power Forecasting}


% \author{Wei Fan}
% \email{weifan@um.edu.mo}
% \affiliation{%
%   \institution{University of Macau}
%   \city{Taipa}
%   \state{Macau SAR}
%   \country{China}
% }

\author{Wei Fan}
\email{weifan.oxford@gmail.com}
\affiliation{%
  \institution{University of Oxford}
  \city{Oxford}
  %\state{Macau SAR}
  \country{UK}
}

\author{Yanjie Fu}
\authornote{Corresponding Authors}
\email{yanjie.fu@asu.edu}
\affiliation{%
  \institution{Arizona State University}
  \city{Tempe}
  \state{Arizona}
  \country{USA}
}

\author{Shun Zheng}
\email{shun.zheng@microsoft.com}
\author{Jiang Bian}
\email{jiang.bian@microsoft.com}
\affiliation{%
  \institution{Microsoft Research}
  \city{Beijing}
  \country{China}
}

\author{Yuanchun Zhou}
\email{zyc@cnic.cn}
\affiliation{
\institution{Computer Network Information Center, Chinese Academy of Sciences}
\city{Beijing}
  \country{China}
}

\author{Hui Xiong}
\authornotemark[1]
\email{xionghui@ust.hk}
\affiliation{
  \institution{Hong Kong University of Science and Technology}
  \city{Guangzhou}
  %\state{Guangzhou}
  \country{China}
}

% \affiliation{%

% \affiliation{%

\renewcommand{\shortauthors}{Fan, et al.}

\begin{abstract}
  Wind is one kind of high-efficient, environmentally-friendly and cost-effective energy source.
Wind power, as one of the largest renewable energy in the world, has been playing a more and more important role in supplying electricity.
Though growing dramatically in recent years, the amount of generated wind power can be directly or latently affected by multiple uncertain factors, such as wind speed, wind direction, temperatures, etc. 
More importantly, there exist very complicated dependencies of the generated power on the latent composition of these multiple time-evolving variables,
which are always ignored by existing works and thus largely hinder the prediction performances.
To this end, we propose \textit{DEWP}, a novel \textit{\underline{D}eep \underline{E}xpansion learning for \underline{W}ind \underline{P}ower forecasting} framework to carefully model the complicated dependencies with adequate expressiveness.
DEWP starts with a stack-by-stack architecture, where each stack is composed of (i) a \textit{variable expansion block} that makes use of convolutional layers to capture dependencies among multiple variables;
(ii) a \textit{time expansion block} that applies Fourier series and backcast/forecast mechanism to learn temporal dependencies in sequential patterns.
These two tailored blocks expand raw inputs into different latent feature spaces which can model different levels of dependencies of time-evolving sequential data.
Moreover, we propose an \textit{inference block} corresponding for each stack, which applies multi-head self-attentions to acquire attentive features and maps expanded latent representations into generated wind power.
In addition, to make DEWP more expressive in handling deep neural architectures, we adapt doubly residue learning to process stack-by-stack outputs. Accurate wind power forecasting is then better achieved through fine-grained outputs by continuously removing stack residues and accumulating useful stack forecasts.
Finally, we present extensive experiments in the real-world wind power forecasting application on two datasets from two different turbines, in order to demonstrate the effectiveness of our approach.
\end{abstract}

\begin{CCSXML}
<ccs2012>
<concept>
<concept_id>10002951.10003227.10003351</concept_id>
<concept_desc>Information systems~Data mining</concept_desc>
<concept_significance>500</concept_significance>
</concept>
</ccs2012>
\end{CCSXML}

\ccsdesc[500]{Information systems~Data mining}

\keywords{Wind Power Forecasting, Time Series Forecasting, Deep Learning}

\maketitle

\section{introduction}
In recent years, the blooming development of the world's economy and population imposes an unprecedented energy crisis that results into negative environmental consequences, such as air pollution, ozone depletion, and global warming \cite{bilal2018wind}.
Renewable energy, the energy that can naturally be replenished on a human timescale, thus becomes a viable solution to overcome the energy crisis while being able to protect the natural environment. 
According to the Paris agreement, renewable energy will make up two-thirds of energy consumption in order to limit global warming to well below two degrees Celsius, compared to pre-industrial levels \cite{gielen2019role}.
%Among all the renewables, wind power has played a more and more important role in supplying electricity, due to its wide availability, high efficiency, and low prices.  There is a large growth for wind power generation around the world: 

Among all the renewables, wind power has played a more and more important role in supplying electricity. The generation of wind power grows very quickly due to its wide availability, high efficiency, and low prices.
For example, in UK, wind power generation has more than doubled from 2016 to 2020 \cite{ukenergy}.
%The UK plans to increase its installed capacity for offshore wind generation to 40 GW by 2030, increasing overall wind capacity to over 50 GW for commitment to achieve net zero carbon emissions by 2050.
Despite the increasing popularity, there is a well known challenge for the operation and management of wind power.
A power grid needs to balance the electricity generation and demand at all times.
%Grid operators must balance the electricity generation and demand on the grid. 
However, wind power generation is uncertain and relates with dynamically changing environmental factors, such as wind speed \cite{yan2015reviews}. 
As wind energy accounts for higher percentage in electricity supply, the uncertainty makes it more difficult to balance power supply and demand.
As a result, high ramping power plants must provide some reserve capacity to meet the demand beforehand \cite{hanifi2020critical}. Hence, it is important to accurately forecast for future electricity generation, in order to manage the integration, maintain reserve capacity, control grid frequency fluctuations, and allocate electricity generation.

Despite its great importance in evaluating future energy extraction, accurate high-resolution wind power forecasting is very challenging as it is affected rapidly by many direct or latent factors.
For example, the speed of wind turbine blades depends on how fast the wind is blowing, and so wind speeds directly influence electricity generation~\cite{hemami2012wind}. 
%higher wind speeds allows the blades to rotate faster, thus making turbine can acquire more power \cite{hemami2012wind}. 
Besides, temperature, humidity or atmospheric pressure indirectly influence wind power generation. 
These factors closely relate to local air density, while different air density asserts different pressure on the rotors. The ``heavier'' the air is, the more energy received by the turbine~\cite{el2017evaluation}.
Other features like the date or the time of a day are also important features for forecasting, as they can reveal distinctive weather conditions like monsoons that can imply wind speed or wind direction. 
Generally, all these factors need to be well-modelled for an accurate forecasting.

\begin{figure}[th]
\centering
\includegraphics[width=0.7\linewidth]{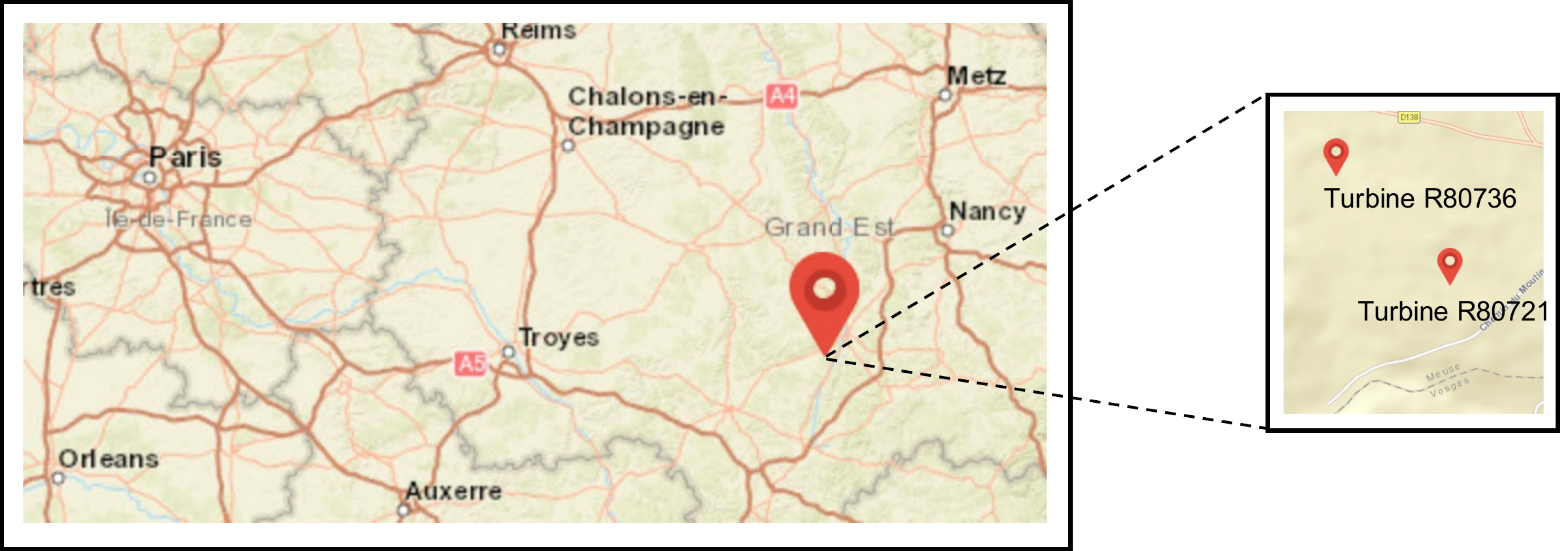}
%\vspace{-2mm}
\caption{Turbine R80736 and Turbine R80721 at `La-Haute-Borne' wind farm.}
\label{fig:map}
%\vspace{-3mm}
\end{figure}

% \begin{figure}[th]
% \centering
% \includegraphics[width=0.7\linewidth]{figure/map.png}
% %\vspace{-3mm}
% \caption{}
% \label{fig:map}
% \vspace{-2mm}
% \end{figure}

\begin{figure}[th]
\centering
\includegraphics[width=0.75\linewidth]{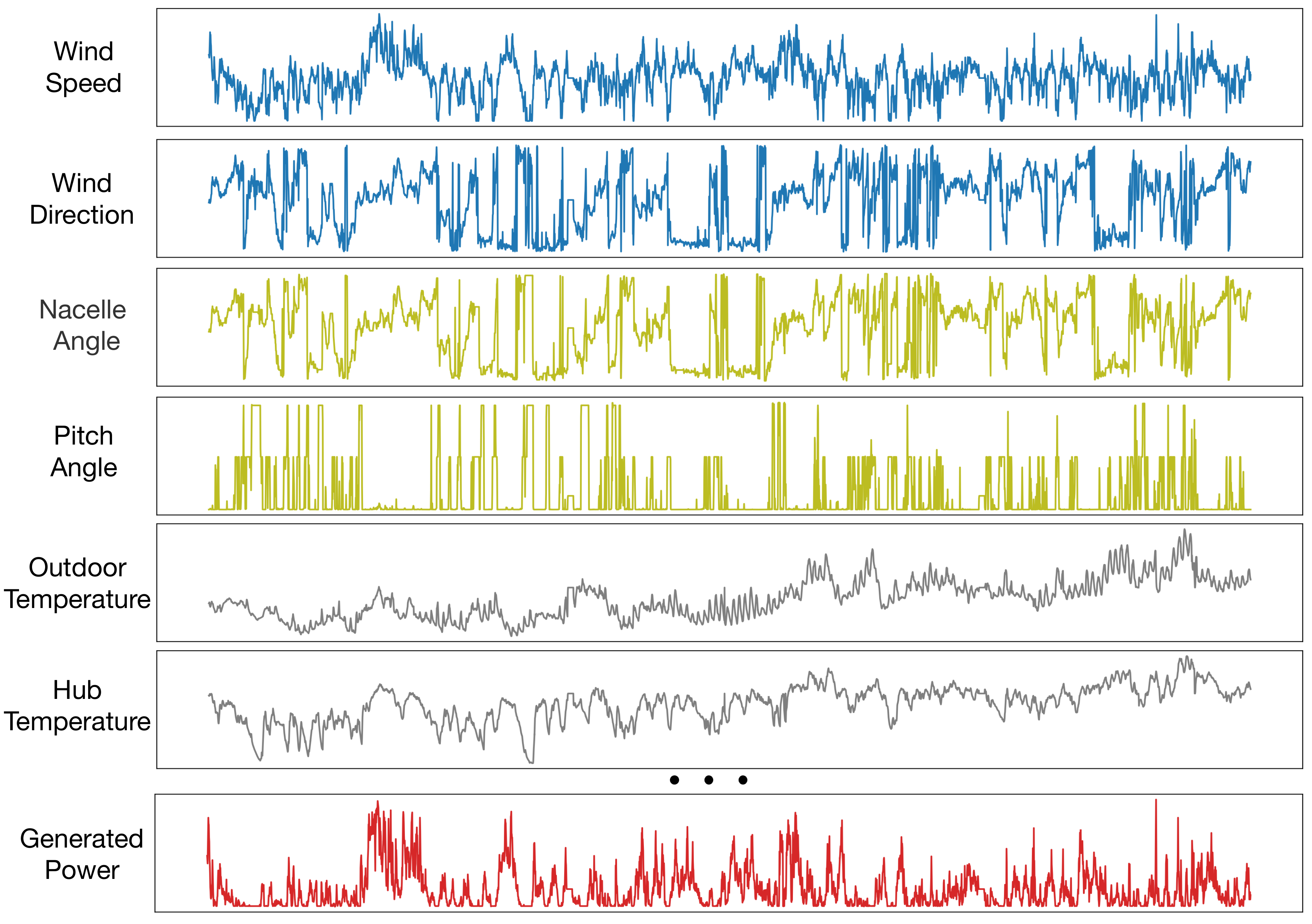}
%\vspace{-2mm}
\caption{An example of several related sequential factor variables and target (generated power) of 6 months since Janurary 1st, 2013 from turbine R80736.}
\label{fig:features}
%\vspace{-3mm}
\end{figure}

\noindent \textbf{Application and Domain Challenges.} 
Our real-world application is using historical natural observations (without power) to predict the amount of wind power generated by the turbines of a wind farm located in north-easter France. 
%near the commune of Vaudeville-le-Haute. 
This 'La-Haute-Borne' farm (Figure \ref{fig:map}) is the first open data wind farm powered by ENGIE \footnote{http://www.engie.com}, which provides electricity to the equivalent of 7,300 people since 2009 thus avoiding around 12,000 metric tons of CO2 emissions per year. 
%The turbines have been in commision since 2009, with the hub height as about 80 meters and rotor diameter as 82 meters. More detailed informations and statistics can be referred to Section \ref{}.
The wind power data are in the form of continuous, real-time, sequential and uniformly sampled data, whose representation is often of multi-variate time series. 
As Figure \ref{fig:features} shows, on the one hand, the raw wind power data includes much uncertainty and noise, and the intermediate fluctuations make the forecasting very difficult. 
On the other hand, the generated power have complicated dependencies on the composition of the multiple time-evolving variables. 
The generated power is latently dependent on the variable-variable correlations (variable dependencies) and time-evolving patterns (temporal dependencies), both of which should be carefully modelled for dependency learning and accurate forecasting.

\noindent \textbf{Our Contributions.} 
Prior studies on wind power forecasting can be grouped into: (i) physical methods that use physical characterisation to model wind turbines/farms \cite{lange2008new}; 
(ii) statistical methods that develop linear or non-linear statistical relationships between observations and generated power \cite{giebel2011state, mahoney2012wind}; 
(iii) hybrid methods combines physical methods and statistical methods together \cite{giebel2006shortterm}. Recently, many deep learning based methods apply CNNs, RNNs and their variants for the forecasting \cite{yu2019lstm, hong2019hybrid}. 
However, most of them don't thoroughly construct dependencies of generated power on intricate sequential input observations
However, most existing works towards wind power forecasting \cite{lange2008new,mahoney2012wind,bilal2018wind,maldonado2021wind} ignore the complicated dependencies of generated power on intricate sequential input variable observations; moreover, they use some simple designs (e.g., CNNs, RNNs) and thus don't have enough network expressiveness to model these dependencies, which largely hinders the prediction performances.

To address the aforementioned challenges, we aim to develop a generic and expressive (deep) architecture to predict the wind power and propose a novel \textit{ \underline{D}eep \underline{E}xpansion learning for \underline{W}ind \underline{P}ower forecasting (DEWP)} framework, which not only carefully model these complicated dependencies but accommodate adequate expressiveness through the network designs.
Specifically, the manually extracted time-evolving variables and the embeddings mapped by timestamps are first taken as raw input to DEWP.
%we first analyze the features and manually extract most important and meaningful variables to compose the dataset. To reduce the noise and randomness, we organize the raw datasets as hourly data and carefully preprocess the data as two benchmarks. Then, we formulate the wind power forecasting problem as a multi-variate point forecasting task, which samples a fixed-length lookback time window of variable as input and a horizon of active wind power as target. 
DEWP starts with a stack-by-stack architecture, which consists of several stacked layers (called \textit{stacks}) to learn dependencies and representations of the input.
%which try to expand the variables into hidden states for feature dependency extraction. 
To learn the complicated dependencies of variable compositions, we propose a \textit{variable expansion block} for each stack, which makes use of convolution layers on variable dimensions to model their latent relationship for feature extraction; the raw input variables are then expanded into hidden states. 
After the variable expansions, DEWP is designed to further capture the continuous temporal dependencies through time expansions in each stack. Accordingly, we propose a \textit{time expansion block} in each stack which accepts variable expanded results and produces expansion coefficients for backcast/forecast mechanism \cite{oreshkin2019n}. In particular, this block further model characteristic of seasonality by adapting Fourier series to constrain the cyclical function, in order to mine the regular recurring fluctuation. These designs empower DEWP to learn time dependencies and carry latent representations to the outputs that mainly consist of two parts: {the backcast part aims to represent historically observed knowledge and the forecast part aims to predict future information towards wind power.}

Moreover, to leverage the latent representations to predict wind power generation, we propose an \textit{inference block}, corresponding for each stack of DEWP, which applies multi-head self-attentions \cite{vaswani2017attention} to process the the forecast part output of the time expansion blocks to acquire attentive features, and map the expanded representations for final forecasting.
In addition, to make DEWP more expressive in handling deep architectures, we follow \cite{oreshkin2019n} to conduct a doubly residue learning from stack to stack: 
the backcast results are continuously removed before taken to the next stack; the output mappings of inference blocks are accumulated to get a fine-grained forecasting towards the final ground truths (generated power).
Finally, we conduct extensive experiments in the real-world application setting on our developed two benchmark datasets, which clearly validate the effectiveness of our approach on the task of wind power forecasting. 
The experimental results demonstrate that DEWP is a good solution for wind power prediction that can provide timely insights to the energy market.

%As discussed in, the two main challenges of wind power forecasting lies in (i) how to capture the sophisticated dependencies of generated wind power on different multiple variables, and (ii) how to capture the continuous dependencies of generated power on time-evolving patterns of sequential data. These two challenges motivate us to develop a generic and expressive (deep) archituctures to capture the mentioned two dependencies.
%Figure \ref{} shows the specific values of several related features/factors as well as the generated power.
%To model complicated dependencies of sequential wind power data, we propose a novel \textbf{D}eep \textbf{E}xpansion learning framework for \textbf{W}ind \textbf{P}ower forecasting, called \textbf{DEWP}.

\section{Preliminaries}

In this section, we will introduce the preliminaries of the real-world wind power forecasting application, including application backgrounds, data description, and problem formulation.

\subsection{Backgrounds and Contexts}
Wind energy is one of the RES characterized by the lowest cost of electricity production and the largest resource available~\cite{wang2011review}. However, the power output from renewable sources such as wind power is highly variable and hard to predict, considering the amount of power coming from renewable sources largely depends on nature, which cannot be thoroughly captured by human beings.
%More importantly, {we are in a way at the behest of nature} when it comes to the problem about how much power we can generate from these sources at any given point of time.

\begin{figure} \centering  
\subfigure[Generated active power curve versus wind speed.] {
 \label{fig:power_windspeed_curve}     
\includegraphics[width=4.5cm]{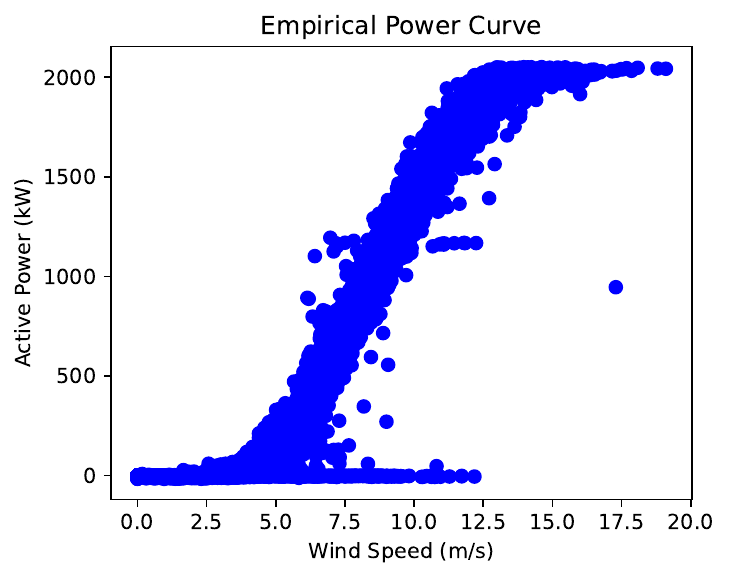} 
}  
\hspace{+4mm}
\subfigure[Wind direction versus wind speed.] { 
\label{fig:wind_angle}     
\includegraphics[width=4cm]{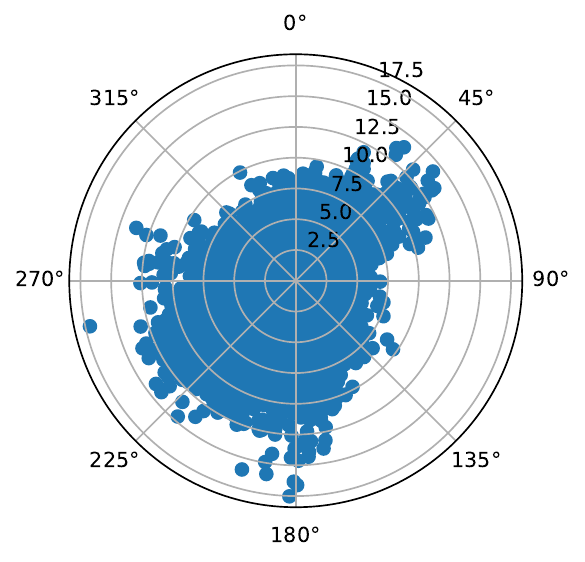}     
} 
\caption{The distribution of wind speed, wind direction, and generated power of a year since Janurary 1st, 2013 from turbine R80736.}     
\label{fig:backgrounds}
\end{figure}

There are many direct or indirect factors that influence wind power generation. 
For example, wind speed largely determines the amount of power generated by a turbine. 
Figure \ref{fig:power_windspeed_curve} shows high-speed winds allow the blades to rotate faster in the range of cut-in speed \footnote{The cut-in speed is the point at which the wind turbine is able to generate power.} and cut-out speed \footnote{The cut-out speed is the point at which the turbine must be shut down to avoid damage to the equipment.}, thus generating more active wind power \cite{hemami2012wind}.
%\footnote{https://energyeducation.ca/encyclopedia/RPM}.
Also, wind direction is another factor that influences power generation. 
Figure \ref{fig:wind_angle} shows the wind directions and wind speeds in 2013
at the location of turbine R80736, and we easily observe that winds from different directions always have different speeds. %When the wind direction is right, 
When the wind reaches a proper direction, the smaller difference between wind angle and the nacelle angle allows for faster rotating speed of turbine; hence, more power is generated.
Some other factors also influence the wind power, such as temperature 
or humidity, which can determine local air density and thus exerts different pressure on the rotors. In brief, the ``heavier'' the air is, the more power is generated by the turbine \cite{el2017evaluation}. Aside from these factors, we need to consider the date or time of a day, because they can imply some distinctive weather conditions such as monsoons. Seasonal climates that influence wind power generation could be represented by these time flags.
%Other features like the date or the time of a day are also important features for forecasting, as they can reveal distinctive weather conditions like monsoons that can imply wind speed or wind direction. 

Moreover, all these factors, which can directly or indirectly influence the final output of wind power, vary over time and have their own latent time-evolving sequential patterns. These multiple variables that can be regarded as multi-variate time series, need to be properly modelled for the accurate forecasting towards dynamic wind power generation.

\subsection{Data Description and Preprocessing} \label{sec_data}

\paragraph{Description.}
We exploit real-world wind power data from ENGIE, the largest power company in France, supplying electricity in 27 countries in Europe and 48 countries worldwide and having its wind-power leadership position in the world. 
The data \footnote{https://opendata-renewables.engie.com/} is collected at the “La Haute Borne” wind farm (in the Meuse department), which provides electricity to the equivalent of 7,300 people since 2009 and reduce around 12,000 metric tons of CO2 emissions per year. 
We extract data of two turbines (R80736 and R80721) to compose two benchmark datasets. 

\paragraph{Preprocessing.}
Since the raw data is collected in every 10 minutes and includes lots of uncertainty and noises, we aggregate the data of both turbines into hourly data to reduce randomness and augment the data patterns. Specifically, the hourly aggregation is completed by averaging all the points available each hour, where the missing values are set as the mean values of the whole dataset. Furthermore, inspired by feature selection process~\cite{fan2020autofs,fan2021interactive}, we manually extract series of meaningful features, including wind speed, wind direction, nacelle angle, pitch angle, and temperatures from different sensors.
Since the start point and the end point of two benchmark datasets are 2013-1-1 and 2016-12-31, we split the data into training set and  test set at the point of 2016 10-01 00:00. The overall timestamps of the datasets are 35069 points, with the first 24076 timestamps belong to the training set. After sliding the time series data into windows, the training data, validation data include 27544 and  3061 data samples, respectively. 
Figure \ref{fig:features} shows an example of our processed multi-variate time series data from Janurary, 2013 to June, 2013 at turbine R80736.
Table \ref{table:stats} shows the specific statistics of the datasets.

\subsection{Problem Formulation}
We study the problem of wind power forecasting with regard to historical time-evolving multiple variables. Due to the complicated dependencies of the generated power on the multi-variate time-evolving sequential data, we formulate the wind power forecasting problem as:

\paragraph{Definition 1.} \textbf{Wind Power Forecasting.} 
We consider the wind power forecasting problem of regularly sampled multi-variate time series. Let $\mathbf{x}_t \in \mathbb{R}^d$ denote the values of multiple series at time-step $t$, where $\mathbf{x}_t = \{ x^1_t, x^2_t, ..., x^d_t \}$ is composed of $d$ scalars each of which represents the value of a variable at $t$-th step. Here each variate stands for a specific feature of wind power: for example, supposing the $i$-th variate $\mathbf{x}^i$ is wind speed, accordingly $\mathbf{x}^i_t$ is the wind speed value (m/s) at time-step $t$. 
Given the historical observations of a certain length $L$, $\mathbf{X}_{t-L:t} = [\mathbf{x}_{t-L}, \mathbf{x}_{t-L+1}, ..., \mathbf{x}_{t-1} ]$, the task of wind power forecasting is to learn a function $\mathcal{F}_\Theta$ to project historical observations to the values of future generated power in a period of time, $\mathbf{y}_{t:t+H}$. 
Formally,
\begin{equation}
    \mathbf{y}_{t:t+H} = \mathcal{F}_\Theta(\mathbf{X}_{t-L:t}) + \epsilon_{t:t+H}
\end{equation}
where $L$ is the length of the lookback window, $H$ is the length of the forecast horizon, $\mathcal{F}_\Theta: \mathbb{R}^L \rightarrow \mathbb{R}^H$ is a mapping function parameterized by $\Theta$, 
and ${\epsilon}_{t:t+H} = [\epsilon_t, \dots, \epsilon_{t+H-1}]$ denotes a vector of independent and identically distributed Gaussian noises.

\begin{table}
\vspace{-0mm}
%\footnotesize
\small
\centering
\caption{Statistics of datasets.}
%\vspace{-3mm}
\begin{tabular}{ccc}
\hline
Datasets & R80736 & R80721 \\
\hline
Turbine GPS& 48.4461, 5.5925 & 48.4497, 5.5869 \\
Rotor Diameter& 82 meters & 82 meters \\ 
\hline
Start Date& 01/01/2013 00&01/01/2013 23 \\
End Date & 12/31/2016 00 & 12/31/2016  23 \\
Time Interval&  Every hour & Every hour \\
\hline
Active Power & [-18.5, 2051.1] & [-17.1, 2051.9]  \\
Wind Speed & [0.0, 20.6]  & [0.0, 19.2] \\
Wind Angel & [0.0, 360.0]  & [0.0, 360.0] \\
Pitch Angle& [-121.2, 119.1] & [-12.4, 120.9] \\
Outdoor Temperature & [-6.3, 37.8]& [-5.9, 38.4] \\
Hub Temperature & [4.0, 37.1] & [0.0, 38.0] \\
%Start Date & 2013-01-01 & 2013-01-01\\
%End Date& 2016-12-31 & 2016-12-31\\
\hline
\end{tabular}
\label{table:stats}
\end{table}

\section{Methodology and Model Design}

For an effective solution for wind power forecasting problem, in this section, we first show an overview of our main framework DEWP and then introduce each component of this framework with more details, including variable expansion block, time expansion block, inference block and expansion residue learning.
% this can write some infos about data processing

% As discussed in, the two main challenges of wind power forecasting lies in (i) how to capture the sophisticated dependencies of generated wind power on different multiple variables, and (ii) how to capture the continuous dependencies of generated power on time-evolving patterns of sequential data. These two challenges motivate us to develop a generic and expressive (deep) archituctures to capture the mentioned two dependencies. As a result, we propose the DEWP Learning Stack, which establishes inherent relationship between the target and the input variables through constructing the Variable Expansion Block (VEB) and Time Expansion Block (TEB). 

\begin{figure*}[th]
\centering
\includegraphics[width=\linewidth]{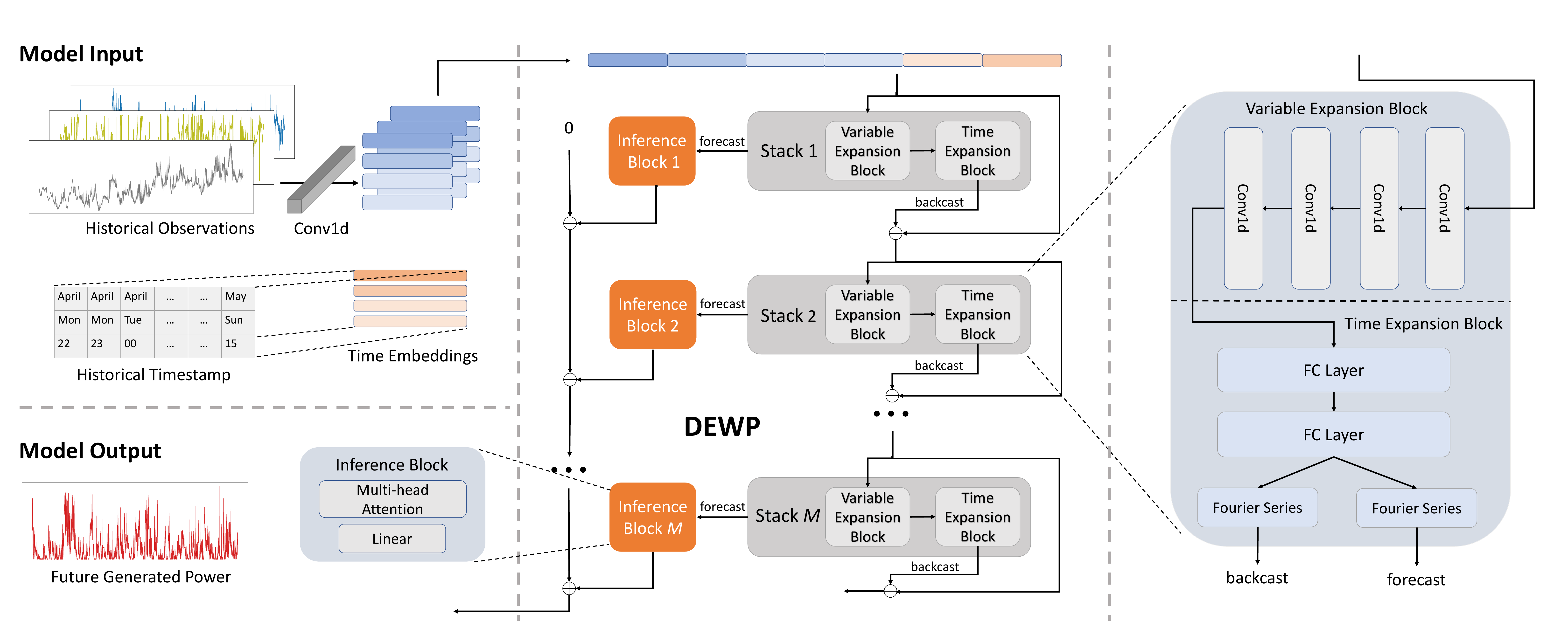}
%\vspace{-6mm}
\caption{Framework Overview. The left part shows the input and output of the framework; the middle part shows the main architecture of DEWP; the right part shows inner components of each stack.}
\label{fig:overview}
\vspace{-0mm}
\end{figure*}

\subsection{Framework Overview}
Figure \ref{fig:overview} shows an overview of our wind power forecasting framework. 
To begin with, the model takes the historical observations of multiple variables (without historical wind power) and timestamp flags (including months flag, weekdays flag and dates flag) as raw input. The input is then processed to convolutional features and time embeddings, and their concatenated vectors are regarded as the input of the main model of DEWP. 
The main model, shown as the middle part of Figure \ref{fig:overview}, is organized as a {stack-by-stack} architecture, where each \textbf{\textit{Stack}} is composed of one \textbf{\textit{Variable Expansion Block}}, and one \textbf{\textit{Time Expansion Block}}. The detailed architecture of each stack and block is shown in right part of Figure \ref{fig:overview}.
The variable/time expansion block aims to capture the complicated dependencies on multiple variables and temporal patterns respectively.
After expansion operations, each stack will carry the hidden expanded states into the outputs that mainly include two parts: the backcast part and the forecast part.
The backcast part, regarded as the stack residues, is subtracted before taken to the next stack, in order to conduct a stack-by-stack residue learning of expansions. 
The forecast part of each stack, is taken to its corresponding \textbf{\textit{Inference Block}} to map the hidden expansion results to future wind power, in order to conduct a stack-by-stack fine-grained inferences. The detailed design of the inference block is shown in bottom left of Figure \ref{fig:overview}.
Finally, the inference results of each inference block are accumulated, in order to finally contribute to the future wind power generation.

\subsection{Time Embeddings}
As Figure \ref{fig:overview} shows, for the input historical observations, apart from extracting hidden embeddings using convolution operations, we also include the time embeddings as the extra temporal information as input to enhance the prediction performance. 
Specifically, we extract three kinds of features as the affiliated information from the timestamps:  
\begin{itemize}
    \item \textit{Month embeddings}: we extract the month information from the timestamp and map different months (e.g., January, February, etc) into different embeddings.
    \item \textit{Weekday embeddings}: we extract the date information from the timestamp and find the according weekday information (e.g., Monday, Tuesday, etc) to generate embddings.
    \item \textit{Hour embeddings}: since the raw data is processed with the hourly aggregation, we thus also extract the hourly information and accordingly generate the hour embeddings.
\end{itemize}

%\vspace{-2mm}
\subsection{Variable Expansion Block}
One of the main challenges of the wind power forecasting lies in how to capture the sophisticated dependencies of the generated wind power on different compositions of multiple variables. The latent relationship between the target and the input variables is non-linear and hard to capture, which motivates us to develop a simple yet effective module for feature extraction and variable dependency learning.

Though convolutional networks have been applied to sequential modelling in many studies \cite{bai2018empirical, borovykh2017conditional, perslev2019u}, all these works need to carefully design covolutions for the chronological order of series. Actually, the classic convolution architecture itself has strong ability of feature extraction, and thus can handle dozens of tasks from different domains \cite{lecun1995convolutional}. In our framework, we propose to apply convolution architectures to model latent variable dependencies for feature extraction.
As the middle part of Figure \ref{fig:overview} shows, the overall framework stacks many variable expansion blocks and time expansion blocks. 
When it comes to the Variable Expansion Block, we want to model latent dependencies only for variables and leave the temporal dependencies to the time expansion block. Such a decoupled dependency learning design motivates us to apply 1-dimensional convolution rather than 2-dimensional convolution. Thus, we adapt 1-dimensional convolution to work on the variable dimension learning for feature extraction.
Specifically, we denote the input to the $\ell$-th stack (the backcast output of $(\ell-1)$ stack) as $\mathbf{X}^{(\ell)}_{t-L:t} \in \mathbb{R}^{d_v*L}$, where $d_v$ is the dimension of hidden states after expansions of previous stacks.
For variable expansion, the input directly accepted by the $\ell$-th stack is processed by four layers of convolution network:
\begin{equation}
\begin{aligned}
   &\mathbf{h}^{(\ell),1}_{t-L:t} =  \sigma(Conv(\mathbf{X}^{(\ell)}_{t-L:t})), \;\;
   \mathbf{h}^{(\ell),2}_{t-L:t} =  \sigma(Conv(\mathbf{h}^{(\ell),1}_{t-L:t})), \\
   &\mathbf{h}^{(\ell),3}_{t-L:t} =  \sigma(Conv(\mathbf{h}^{(\ell),2}_{t-L:t})), \;\;
   \mathbf{z}^{(\ell)}_{t-L:t} =  Conv(\mathbf{h}^{(\ell),3}_{t-L:t})
\end{aligned}
\end{equation}
where $Conv$ stands for the standard 1-$d$ convolution operations \cite{o2015introduction} and $\sigma$ is the ReLU non-linearity \cite{glorot2011deep}.
This variable expansion block expands the input variables $\mathbf{X}^{(\ell)}_{t-L:t}$ with convolutions and emits the expanded features $\mathbf{z}^{(\ell)}_{t-L:t} \in \mathbb{R}^{d_v*L}$, which is taken as input to time expansion block of $\ell$-th stack.

%\subsubsection{Variable Expansion Block}
%The variable expansion blocks accepts the input to the l-th learning stack $x_l, ....$, which is the hidden expansion basis after expericing the expansion of block 1 to l-1. Note that we aim to distill the hidden features of input variables, we adopt 1-d convolutional operations to extract mixed features of input.
\vspace{-2mm}
\subsection{Time Expansion Block}
Aside from dependencies among variables, another main challenge of wind power forecasting is how to capture the continuous dependencies of generated power on time-evolving patterns of sequential data. After representation based on variable expansions, we also need to accurately model temporal dependencies for future values based on the historical observations of time series.

Many studies have tried different strategies for sequential data, such as recurrent neural networks \cite{gers2002applying} or attention \cite{vaswani2017attention}. However, most of them suffer from large computational burden. 
Recently, it has shown great success of pure fully-connected networks for temporal basis expansion of series in order for accurate forecasting \cite{oreshkin2019n,fan2022depts}. On top of them, we develop a time expansion block with backcast/forecast mechanism \cite{oreshkin2019n} for deep expansion learning across time steps. 
First, the $\ell$-th time expansion block will accept $\ell$-th variable expansion results $\mathbf{z}^{(\ell)}_{t-L:t}$ as input and process it with two fully-connected layers with ReLU non-linearity:
\begin{equation}
\begin{aligned}
  &\mathbf{z}^{(\ell),1}_{t-L:t} = \sigma(\mathbf{w_{z}} \mathbf{z}^{(\ell)}_{t-L:t} + \mathbf{b_{z}})\\
  &\mathbf{\rho}^{(\ell)} = \sigma(\mathbf{w_{\rho}} \mathbf{z}^{(\ell),1}_{t-L:t}+\mathbf{b_\rho})
 \end{aligned}
\end{equation}
where the projected results $\mathbf{\rho}^{(\ell)} \in \mathbb{R}^{d_v*(L+H)}$ is called expansion coefficients of length $L+H$, for the following backcast and forecast. 
Specifically, we let the backcast part recover the historical observation signals ( $[t-L, t]$ steps) through prediction, while let the forecast part is to predict future signal values ($[t, t+H]$ steps).
%With the expansion coefficients, we further consider the characteristic of seasonality and adapt the seasonality layer with a backcast/forecast mechanism \cite{} for the time expansion.
Thus, we define two time vectors on a discrete grid to flag the two parts: 
the backcast time vector $\mathbf{t_b} = [-L, -L+1, ..., 0]^T/(L+H)$, 
and the forecast time vector $\mathbf{t_f} = [0, 1, ..., H]^T/(L+H)$.
%With the expansion coefficients, we  further consider the characteristic of seasonality and adapt a seasonality layer \cite{} for time expansion.

For further consideration of characteristic of seasonality, we use Fourier series to constrain the cyclical function in order to capture the regular recurring fluctuation. Accordingly, we can the get backcast expansion results with the time vector:
\begin{equation}
\widehat{\mathbf{X}}_b^{(\ell)}=\sum_{i=0}^{\lfloor L / 2-1\rfloor} 
\rho^{(\ell)}_{i} \cos (2 \pi i)\mathbf{t_b} +
\rho^{(\ell)}_{i+\lfloor L / 2\rfloor} \sin(2 \pi i)\mathbf{t_b}
%\theta_{s, \ell, i+\lfloor H / 2\rfloor}^{f} \sin (2 \pi i t)
\end{equation}
where $\rho^{(\ell)}_i$ is the $i$-th dimension of $\rho^{(\ell)}$ and the first $L$ dimensions are corresponding to the backcast.
Symmetrically, we continue using Fourier series to constrain expansion coefficients and then get forecast expansion results with the time vector:
\begin{equation}
\widehat{\mathbf{X}}_f^{(\ell)}=\sum_{i=L}^{\lfloor L+ H / 2-1\rfloor} 
\rho^{(\ell)}_{i} \cos (2 \pi i)\mathbf{t_f} +
\rho^{(\ell)}_{i+\lfloor H / 2\rfloor} \sin(2 \pi i)\mathbf{t_f}
%\theta_{s, \ell, i+\lfloor H / 2\rfloor}^{f} \sin (2 \pi i t)
\end{equation}
where the last $H$ dimensions of $\rho^{(\ell)}$ are corresponding for the forecast.
The time expansion block expand observations to model the temporal dependencies through backcast and forecast; thus time-evolving patterns can be better captured.
Afterwards, the backcast results $\widehat{\mathbf{X}}_b^{(\ell)} \in \mathbb{R}^{d_v*L}$ will be taken to the $\ell+1$ stack; the forecast results $\widehat{\mathbf{X}}_f^{(\ell)} \in \mathbb{R}^{d_v*H}$ will be input to the corresponding inference block for the final wind power forecasting. 
More details are illustrated in the following sections.

\subsection{Inference Block}
The aforementioned two expansion architecture extract features and address the dependency challenges from two different levels. After getting the expanded representations, the next challenge of wind power forecasting is how to accurately map the hidden expansion results to the actual generated power. For this problem, we design an inference block corresponding for each variable/time expansion block to conduct stack-by-stack fine-grained inferences. 
% Self-attention An attention function can be described as mapping a query and a set of key-value pairs to an output

Specifically, given forecast expansion results $\widehat{\mathbf{X}}_f^{(\ell)}$, we aim to apply the multi-ahead attention architectures \cite{vaswani2017attention} to acquire the attentive features, where the standard scaled dot-product attention is used, written by:
\begin{equation}
\operatorname{Attention}(Q, K, V)=\operatorname{softmax}\left(\frac{Q K^{T}}{\sqrt{d_{k}}}\right) V
\end{equation}
where $Q, K, V$ stands for the queries, keys and values vector and $d_k$ is the dimensions of keys. 
To compose query vectors, key vectors and value vectors, $\widehat{\mathbf{X}}_f^{(\ell)}$ is firstly mapped by three fully-connected networks into $\hat{\mathbf{X}}^{(\ell)}_{f,Q}, \hat{\mathbf{X}}^{(\ell)}_{f,K}, \hat{\mathbf{X}}^{(\ell)}_{f,V}$.
Then, we let $Q, K, V$ equal to $\hat{\mathbf{X}}^{(\ell)}_{f,Q}, \hat{\mathbf{X}}^{(\ell)}_{f,K}, \hat{\mathbf{X}}^{(\ell)}_{f,V}$ respectively; note that the self-attention is on first dimension of $\widehat{\mathbf{X}}_f^{(\ell)}$ (the variable-expanded dimension) to get attentive features. Finally, a linear layer is used to project the attentive features to the target power values. Formally, the whole process is written by:
\begin{equation}
    \hat{\mathbf{X}}^{(\ell)}_{f,Q} = Linear_Q (\hat{\mathbf{X}}^{(\ell)}_{f}),\;
    \hat{\mathbf{X}}^{(\ell)}_{f,K} = Linear_K (\hat{\mathbf{X}}^{(\ell)}_{f}), \;
    \hat{\mathbf{X}}^{(\ell)}_{f,V} = Linear_V (\hat{\mathbf{X}}^{(\ell)}_{f})
\end{equation}
\begin{equation}
    \hat{\mathbf{y}}^{(\ell)}_{f} = Linear_{O}(\operatorname{Attention}(\widehat{\mathbf{X}}_{f,Q}^{(\ell)}, \widehat{\mathbf{X}}_{f,K}^{(\ell)}, \widehat{\mathbf{X}}_{f,V}^{(\ell)})^T)
\end{equation}
where $Linear_Q, Linear_K, Linear_V$ are standard linear layers to project query, key, value vectors; $Linear_O$ is the output linear projection; $\hat{\mathbf{y}}^{(\ell)}_{f} \in \mathbb{R}^H$ is the projected wind power of $\ell$-th stack.

\subsection{Expansion Residue Learning}
Both the variable dependencies and temporal dependencies are well modelled by two expansion blocks, and the inference block effectively constructs mappings from the forecast states to the empirical power. 
To further build a expressive and generalizable model for wind power forecasting, we follow previous work \cite{oreshkin2019n} to conduct a doubly residue learning for further expansions. 

The residue connections include two branches: 
one branch corresponds to the backcast, which helps expansion blocks profoundly represent the existing knowledge from historical observations. Accordingly, for the $\ell$-th stack backcast output $\widehat{\mathbf{X}}_b^{(\ell)}$, we remove it to compose the input to the next stack. As a result, the input to stack $\ell+1$ is given by:
\begin{equation}
    \mathbf{X}^{(\ell+1)}_{t-L:t} = \mathbf{X}^{(\ell)}_{t-L:t} - \widehat{\mathbf{X}}_b^{(\ell)}
\end{equation}
where the $\mathbf{X}^{(\ell)+1}_{t-L:t}, \mathbf{X}^{(\ell)}_{t-L:t}$ is the input to stack $\ell+1$ and stack $\ell$. The other branch corresponds to the forecast, which accumulates output of each inference block for the final predictions. The doubly residue learning here  makes each block learn the residues from layer to layer (backcast), which can be seen as the expansion (the decomposition) for the input signal; thus we need the composition for the output signal (forecasting results).
As a result, the final predictions $\hat{\mathbf{y}}_{t:t+H}$ are:
\begin{equation}
    \hat{\mathbf{y}}_{t:t+H} = \sum_{\ell=0}^{M} \hat{\mathbf{y}}^{(\ell)}_{f}
\end{equation}
where $M$ is the number of inference blocks.
Finally, the main optimization goal of our framework is to minimize the mean-square error of predictions and ground truths, which can be written by:
\begin{equation}
    \mathcal{L} = \frac{1}{N} \sum_{i=1}^{N} \sum_{T=t}^{t+H} (y_T - \hat{y_T})^2 
\end{equation}
where $\mathcal{L}$ is the loss function to minimize, $N$ is the number of input batches, and $y_T, \hat{y_T}$ is the true values and predicted values at $T$-th time step.

\subsection{Model Discussions}
The main superiority of DEWP lies in {\textit{the stacked multi-view expansion architecture}} in summary.
The stacked multi-view expansion architecture of DEWP brings up two important merits:
\begin{itemize}
    \item Multi-view Dependency Modeling Ability of the Expansion Mechanism. In DEWP, we propose variable expansion blocks to capture dependencies among multiple variables and time expansion blocks with apply Fourier series and backcast/forecast mechanism to learn temporal dependencies among multiple timestamps. Such two expansions are iteratively and interactively operated on the raw wind power data, which can capture the complicated dependencies from different aspects (among variables and timestamps). Though existing methods also capture some dependencies, their modeling ability are limited: for example, some methods \cite{yuan2017wind,xie2018nonparametric} usually neglected the dependencies among variables; some works adopt recurrent units of Recurrent Neural Networks but they can only learn dependencies sequentially from single direction~\cite{kisvari2021wind}; however, our temporal expansion blocks are built upon multi-layer perceptrons and Fourier series which capture bi-directional fully connected dependencies.
    \item Expressive Learning Ability of the Stacked Network Architecture. In DEWP, to cooperate different expansion operations of blocks, we set the main framework as a stack-by-stack architecture and utilize doubly expansion residual learning to couple the stacks and blocks. Such a stacked and residue architecture brings much more complicity of networks and enlarge their expressiveness \cite{he2016deep,oreshkin2019n}. Compared with existing wind power forecasting methods, those shallow network designs (CNNs, RNNs) may easily reach to the bottleneck of learning ability. Though vanilla Transformers \cite{vaswani2017attention} with adequate expressiveness can be applied into wind power forecasting, their computation complexity is extremely high, which is not fit for the training, debugging and deployment.
\end{itemize}

\section{Experiments}
We conducted extensive experiments to validate the performance of DEWP framework on two real-world datasets.

%We take real-world wind power data from ENGIE, as the largest power company in France, supplying electricity in 27 countries in Europe and 48 countries worldwide and and having its wind-power leadership position in the world.  The data is collected at “La Haute Borne” wind farm (in the Meuse department), which provids electricity to the equivalent of 7,300 people since 2009, thus avoiding around 12,000 metric tons of CO2 emissions per year\footnote{https://opendata-renewables.engie.com/}. We extract data of two turbines, R80736 and R80721, to compose our two datasets. To reduce randomness and uncertainty of raw data, both datasets are aggregated to hourly data. Since the raw data is sampled every 10 minutes,  the hourly aggregation is done by averaging all the points available each hour, where the missing values are set as the mean values of the whole data. The extracted features include wind speed, wind direction, nacelle angle, pitch angle, and different temperatures. Figure \ref{fig:features} shows an example of multiple time series from Janurary, 2013 to June, 2013. More specific statistic information of datasets are shown in Table 1.

%\vspace{-1mm}
\subsection{Experimental Setup}
\subsubsection{Datasets}
%To compare the performance of different methods, we conduct experiments on real-world data from two turbines at “La Haute Borne” wind farm. 
As illustrated in Section \ref{sec_data}, our experiments were conducted on real-world data of two turbines at “La Haute Borne” wind farm. We developed two benchmark datasets \textit{\textbf{R80736}} and \textit{\textbf{R80721}} by carefully preprocessing the raw data in following steps: 
(i) segmenting the timestamps to align with the calendar,
(ii) completing the missing points or invalid points with mean values, 
(iii) aggregating the raw samples to hourly data samples, 
(iv) conducting feature selection manually on given raw features, and 
(v) utilizing the Min-Max normalization to regularize the raw values into a range of [0,1].
More details of datasets are in Section \ref{sec_data}.

\subsubsection{Metrics.}
To compare different models, we utilized three different widely-used metrics of time series prediction for evaluation: 
\noindent 1) Mean Absolute Error ($MAE=\frac{1}{n} \sum_{i=1}^{n}|y_i - \hat{y}_i|$), 
\noindent  2) Mean Absolute Percentage Error ($MAPE=\sum_{i=1}^{n}|y_i - \hat{y}_i|/|y_i|$), 
\noindent  3) Mean Squared Prediction Error ($MSPE = \sum_{i=1}^{n}(y_i - \hat{y}_i)^2/|y_i|$), 
\noindent where $n$ is the number of points for evaluation, $y_i$ and $\hat{y}_i$ are the target value (generated power) and the predicted value of $i$-th point.

\subsubsection{Evaluation.} It is widely recognized that wind power forecasting can be classified into short/long-term forecasting based on the forecasting horizon \cite{wang2011review}. In the experiments, our evaluations were conducted with the horizon as 24 hours ahead (\textsc{Short-Term}) and 48 hours ahead (\textsc{Long-Term}) following the definitions of previous work \cite{wang2011review}.
We split the datasets at the time point of ``2016-07-01 00:00'' into training data and test data. To compare fairly, we used the rolling-based strategy to predict the future wind power from `2016-07-01 00:00'' to `2016-12-31 23:50'', where the rolling interval is set the same as the size of horizon, in order to perfectly match the length of test data. The metrics were then calculated on the forecasting values and targets.

\subsubsection{Implication Details}
Our model and all the deep learning baselines were implemented with PyTorch. All methods were evaluated on a Linux server with one RTX 3090 GPU. For a robust evaluation, we run all the models for three times in fixed random seeds.
We averagely sampled training data and used 10\% as the validation set. 
The number of stacks $M$ was set as 5. The variable expansion block was set as  4 convolutional layers with hidden channels as 128. For time expansion block, the layer size (the hidden state dimensions $d_v$) was set as 512. The inference block was with 8 heads self-attention and 1 linear layer. 
For training, we set the batch size as 256 and used the Adam Optimizer with a learning rate of 1e-4 with proper early stopping.
More hyper-parameter analysis are in Section \ref{sec:parameters}.

\begin{table*}[t]
\footnotesize
\centering
\caption{Overall performance comparisons of \textsc{Short-Term} wind power forecasting on R80736 dataset.}
\begin{tabular}{c|ccc|ccc}
%\hline \textbf{Datasets} & \multicolumn{6}{|c}{ \textbf{R80736} } \\
\hline \textbf{Lookback} & \multicolumn{3}{|c|}{ \textbf{24} } & \multicolumn{3}{c}{ \textbf{72} } \\
\hline \diagbox{\textbf{Method}}{\textbf{Metrics}} & \textbf{MAPE} & \textbf{MSPE} & \textbf{MAE} & \textbf{MAPE} & \textbf{MSPE} & \textbf{MAE}  \\
\hline
Linear & 5.643 & 89.11 & 0.7596 & 3.947 & 46.74 & 0.6050     \\
GRU    & 3.315 & 35.29 & 0.5528 & 3.371 & 36.62 & 0.5613    \\
BiLSTM& 3.793 & 43.29 & 0.5936 & 3.449 & 37.47 & 0.5651  \\
LSTNet& 3.594&42.19&0.5823& 3.281&35.98&0.5608 \\
TCN &  3.568&  39.26&0.5692 & 3.235&35.76&0.5592 \\
WPF-GRN & 3.303 & 34.81 & 0.5518 & 3.231 & 34.13 & 0.5591 \\
WPF-TSA & 3.288 & 34.32 & 0.5501 & 3.226 & 33.62 & 0.5513 \\
Transformer& 3.645 & 41.82 & 0.5805 &  3.339 &  36.94  & 0.5675     \\
Autoformer&6.506 & 146.0 & 0.9785& 6.364& 138.9 &1.0192     \\
Informer & 3.503 & 38.77 &  0.5733 & 3.110& 31.65&0.5490  \\
N-BEATS & 3.292&34.89&0.5503 & 3.210&32.31&0.5520      \\
\hline 
\textbf{DEWP}   & \textbf{2.848} & \textbf{28.12} & \textbf{0.5220} & \textbf{2.973} & \textbf{28.03} & \textbf{0.5403} \\
\hline
\end{tabular}
\label{tab:overall_s_1}
\vspace{-1mm}
\end{table*}

\begin{table*}[t]
\footnotesize
\centering
\caption{Overall performance comparisons of \textsc{Short-Term} wind power forecasting on R80721 dataset.}
\begin{tabular}{c|ccc|ccc}
%\hline \textbf{Datasets} & \multicolumn{6}{|c}{ \textbf{R80721} } \\
\hline \textbf{Lookback} & \multicolumn{3}{|c|}{ \textbf{24} } & \multicolumn{3}{c}{ \textbf{72} } \\
\hline \diagbox{\textbf{Method}}{\textbf{Metrics}} & \textbf{MAPE} & \textbf{MSPE} & \textbf{MAE} & \textbf{MAPE} & \textbf{MSPE} & \textbf{MAE}  \\
\hline
Linear  & 3.936 & 50.40 & 0.5557 & 5.513 & 90.73 & 0.6923     \\
GRU    & 3.451 & 41.78 & \textbf{0.5163} & 3.817 & 47.85 & 0.5431     \\
BiLSTM & 3.612 & 44.69 & 0.5265 & 4.018 & 52.75 & 0.5564    \\
LSTNet& 3.581&43.89&0.5463 & 3.995&50.08&0.5474\\
TCN &  3.416&42.64&0.5312 &3.851&47.31&0.5412\\
WPF-GRN & 3.605 & 44.42 & 0.5234 & 3.821 & 45.13 & 0.5403 \\
WPF-TSA & 3.594 & 43.12 & 0.5203 & 3.796 & 44.92 & 0.5348 \\
Transformer&  4.051& 53.06 & 0.5584& 4.010 & 54.24 & 0.5502     \\
Autoformer& 4.441 & 69.39&0.6827&  4.162 & 72.54 & 0.7010           \\
Informer &  3.597&43.25& 0.5203 & 3.613 & 40.96&0.5288            \\
N-BEATS & 3.589&42.98&0.5198&  3.721&44.76&0.5398          \\
\hline 
\textbf{DEWP}  & \textbf{3.326} & \textbf{37.25} & 0.5185 &
\textbf{3.436} & \textbf{38.04}  & \textbf{0.5178} \\
\hline
\end{tabular}
\label{tab:overall_s_2}
\vspace{-2mm}
\end{table*}

\subsubsection{Baseline Algorithms}
To comprehensively validate the performance of DEWP framework, we compared it with several representative baseline methods, which mainly include:
\begin{itemize}
    \item \textbf{Linear}: is a standard benchmark model that consists two fully-connected layers with ReLU activations \cite{glorot2011deep} to directly map the historical observations to future generated power.
    \item \textbf{GRU}: is a variant of Recurrent Neural Network (RNN) constrained by gate control, which is widely adopted in time series tasks and wind power forecasting \cite{li2019short}.
    \item \textbf{BiLSTM}: is another RNN-based variants, capturing the information of long-short term memory and being improved by concatenating the forward and backward hidden states.
    \item \textbf{LSTNet} \cite{lai2018modeling}: is a CNN+RNN time series modeling method, which can captures the short-term local dependency patterns and long-term time series trend patterns with covolution and recurrent neural network.
    \item \textbf{TCN} \cite{bai2018empirical}: is a generic temporal convolutional network architecture tailored-designed for sequential modelling, including wind power forecasting \cite{gan2021temporal}.
    \item \textbf{WPF-GRN} \cite{kisvari2021wind}: is a data-driven method specifically designed for wind power forecasting including data pre-processing, re-sampling, feature engineering process and the gated recurrent units as the network for data driven learning.
    \item \textbf{WPF-TSA} \cite{tian2022developing}: is a two-stage attention based forecasting methods, including a feature decomposition module to remove superfluous noise and two-stage attention module to determine the importance of wind features.
    \item \textbf{Transformer} \cite{vaswani2017attention}: is a sequence modelling method that introduces self-attention based architecture for processing multi-variate time-series data.
    \item \textbf{Autoformer} \cite{xu2021autoformer}: is a novel decomposition architecture with an Auto-Correlation mechanism by conducting the dependency learning and representation aggregation for series.
    \item \textbf{Informer} \cite{zhou2021informer}: is a variant of transformer which improves transformer in time complexity, high memory usage, and limitation of the encoder-decoder architecture.
    \item \textbf{N-BEATS} \cite{oreshkin2019n}: is a deep neural architecture for time series forecasting with pure fully-connected networks. Since its original version is for uni-variate time series we modify N-BEATS into a multi-variate version.
\end{itemize}
%All these models are representative deep learning methods showing much effectiveness in time series forecasting. Note that we don't include traditional statistical methods (e.g., ARIMA) as baselines. 

\begin{table*}[t]
\footnotesize
\centering
\caption{Overall performance comparisons of \textsc{Long-Term} wind power forecasting on R80736 dataset.}
\begin{tabular}{c|ccc|ccc}
%\hline \textbf{Datasets} & \multicolumn{6}{|c}{ \textbf{R80736} } \\
\hline \textbf{Lookback} & \multicolumn{3}{|c|}{ \textbf{48} } & \multicolumn{3}{c}{ \textbf{72} } \\
\hline \diagbox{\textbf{Metrics}}{\textbf{Method}} & \textbf{MAPE} & \textbf{MSPE} & \textbf{MAE} & \textbf{MAPE} & \textbf{MSPE} & \textbf{MAE}  \\
\hline
Linear& 4.734&  65.52&  0.6747 &  4.410&  56.05& 0.6430     \\
GRU    & 4.292 & 52.29 & 0.6342 & 4.184 & 52.03 & 0.6204     \\
BiLSTM& 4.735 & 64.75 & 0.6750 & 4.472 & 57.47 & 0.6449  \\
LSTNet& 4.481&55.28&0.6581& 4.397 & 54.38&0.6391 \\
TCN &   4.232&51.70&0.6291& 4.210 & 52.37 & 0.6269  \\
WPF-GRN & 4.112 & 50.33 & 0.6328 & 4.162 & 51.34 & 0.6291 \\
WPF-TSA & 3.987 & 48.23 & 0.6131 & 4.007 & 51.16 & 0.6131 \\
Transformer&  4.270 & 53.58 & 0.6311 &  4.763 & 66.68 & 0.6696     \\
Autoformer& 6.740 & 141.0 & 1.0338 &  4.519 &73.23&0.8983  \\
Informer &  3.920&47.63&0.6167 & 4.193&52.51& 0.6274   \\
N-BEATS & 3.933&46.91&0.6230& 4.186&51.71& 0.6221             \\
\hline 
\textbf{DEWP}   & \textbf{3.876} & \textbf{42.75} & \textbf{0.6083} & \textbf{3.976} & \textbf{44.70} & \textbf{0.6144}\\
\hline
\end{tabular}
\label{tab:overall_l_1}
\end{table*}

\begin{table*}[t]
\footnotesize
\centering
\caption{Overall performance comparisons of \textsc{Long-Term} wind power forecasting on R80721 dataset.}
\begin{tabular}{c|ccc|ccc}
%\hline \textbf{Datasets} & \multicolumn{6}{|c}{ \textbf{R80721} } \\
\hline \textbf{Lookback} & \multicolumn{3}{|c|}{ \textbf{48} } & \multicolumn{3}{c}{ \textbf{72} } \\
\hline \diagbox{\textbf{Method}}{\textbf{Metrics}} & \textbf{MAPE} & \textbf{MSPE} & \textbf{MAE} & \textbf{MAPE} & \textbf{MSPE} & \textbf{MAE}  \\
\hline
Linear&  5.645 & 90.73 & 0.6952& 5.111&76.62&0.6629     \\
GRU    & 5.062 & 75.39 & 0.6361 & 4.492 & 59.93 & 0.5903  \\
BiLSTM&  5.017 & 74.42 & 0.6336 & 4.898 & 72.45 & 0.6201   \\
LSTNet&  4.965&74.13&0.6293&4.601&66.21&0.6038\\
TCN &    4.831& 75.08 & 0.6251 &  4.423&58.96&0.5882          \\
WPF-GRN & 4.839 & 74.63 & 0.6178 & 4.492 & 60.33 & 0.5903 \\
WPF-TSA & 4.781 & 77.12 & 0.6201 & 4.526 & 59.92 & 0.5783 \\
Transformer&  4.917 & 75.52 & 0.6131& 5.007 & 79.10 &  0.6203 \\
Autoformer& 4.972&78.91&0.6728&  5.478 &    75.31 &  0.6835      \\
Informer &  4.759& 77.12 & 0.6222 &  4.411 & 57.29 & 0.5719               \\
N-BEATS & 4.733&73.79&0.6105& 4.506&58.41&0.5841             \\
\hline 
\textbf{DEWP}  & \textbf{4.379} & \textbf{58.03} & \textbf{0.5801} &
\textbf{4.126} & \textbf{54.05}  & \textbf{0.5702} \\
\hline
\end{tabular}
\label{tab:overall_l_2}
\end{table*}

\subsection{Overall Performances}
Table \ref{tab:overall_s_1} and Table \ref{tab:overall_s_2} shows the evaluation of all methods in the \textsc{Short-Term} wind power forecasting task on two datasets.
We observed that \textit{DEWP} beats all the baseline algorithms and achieve the smallest prediction error in most cases, which demonstrates the superiority in short-term wind power forecasting. For example, \textit{DEWP} improves MAPE of the second best result about $(3.292-2.848)/3.315 \approx 13.4\%$ in R80736 dataset. The improvement of MASE in R80721 dataset also comes to $(41.78-37.25)/41.78 \approx 10.8\%$.
We found that  \textit{DEWP} with lookback as 24 can achieve the competitive performance of lookback as 72. This verifies the strong prediction ability even with short lookbacks of our method.
We observed \textit{Linear} model always has the worst performances compared with other methods. This reveals the relationship between future generated power and historical observations is a latent non-linear mapping. The superiority of DEWP with regard to other forecasting methods also indicates that the baseline methods don't have enough expressiveness and don't model complicated dependencies.
Thus we need the delicate modelling of deep learning designs such as \textit{DEWP}.

Table \ref{tab:overall_l_1} and Table \ref{tab:overall_l_2} shows the performance comparisons of the \textsc{Long-Term} wind power forecasting on two datasets. Firstly, we  observed the overall performances are worse than short-term forecasting. This is intuitive considering the historical information is not enough to predict the distant future points, as wind power includes too much uncertainty. In this case, we can still observe \textit{DEWP} outperforms all other baselines and achieves the best performances in all metrics, which signifies better ability for long-term wind power forecasting.
Among the baselines, attention-based methods (e.g., \textit{informer}) usually have better performances than RNN-based methods (e.g., \textit{LSTNet}, \textit{WPF-GRN}), which demonstrates long-term dependencies can be better learned by self-attention layers.
\textit{DEWP} also makes use of self-attention for inference, which helps long-term forecasting.

% \begin{figure*} \centering  
% \subfigure[\small R80736] {
% \label{pa_exp1a}     
% \includegraphics[width=4cm]{exp_fig/pa_lookback_0.pdf} 
% }  
% %\hspace{-2mm}
% \subfigure[\small R80721] {
%  \label{pa_exp1b}     
% \includegraphics[width=4cm]{exp_fig/pa_lookback_1.pdf} 
% }  
% \subfigure[\small R80736] {
% \label{pa_exp2a}     
% \includegraphics[width=4.cm]{exp_fig/pa_horizon_0.pdf}     
% }
% %\hspace{-2mm}
% \subfigure[\small R80721] {
% \label{pa_exp2b}     
% \includegraphics[width=4.cm]{exp_fig/pa_horizon_1.pdf}     
% }
% \subfigure[\small R80736] { 
% \label{pa_exp3a}     
% \includegraphics[width=4.cm]{exp_fig/pa_stack_0.pdf}     
% }
% \subfigure[\small R80721] { 
% \label{pa_exp3b}     
% \includegraphics[width=4.cm]{exp_fig/pa_stack_1.pdf}     
% }
% %\hspace{-2mm}
% \subfigure[\small R80736] { 
% \label{pa_exp4a}     
% \includegraphics[width=4.cm]{exp_fig/pa_layer_0.pdf}     
% }
% %\hspace{-2mm}
% \subfigure[\small R80721] { 
% \label{pa_exp4b}     
% \includegraphics[width=4.cm]{exp_fig/pa_layer_1.pdf}     
% }

% \caption{Parameter analysis towards lookback, horizon, stack and layer size on R80736 and R80721 datasets.}     
% %\label{exp1}

% \end{figure*}

\subsection{Ablation Tests}
To further verify our critical designs of \textit{DEWP}, we conducted extensive ablation tests under different setups. 
We considered four ablated variants of \textit{DEWP} corresponding for each component. Specifically, they are
(i) \textit{DEWP-V}: removing the variable expansion blocks from all the stacks where each stack degenerates into a time expansion block.
(ii) \textit{DEWP-T}: removing the time expansion blocks from all the stacks where each stack degenerates into a variable expansion block.
(iii) \textit{DEWP-I}: replacing the inference block of the original model with a single fully-connected layer for mapping.
(iv) \textit{DEWP-R}: removing all the residue connections of the original model and directly taking outputs of the last stack as the forecasting results.

\begin{figure*} \centering  
\subfigure[\small lookback 24, R80736] {
%\label{pa_exp1a}     
\includegraphics[width=3.2cm]{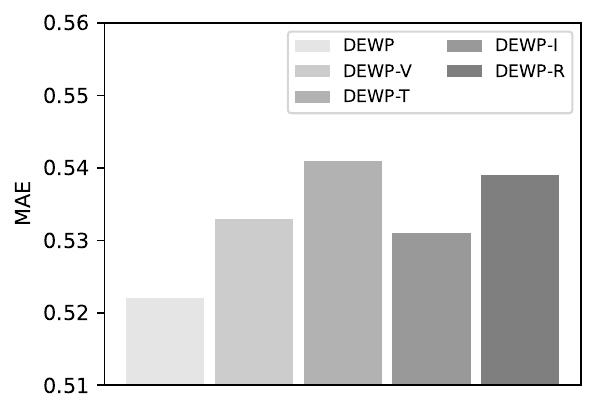} 
}  
%\hspace{-2mm}
\subfigure[\small lookback 72, R80736] {
   
\includegraphics[width=3.2cm]{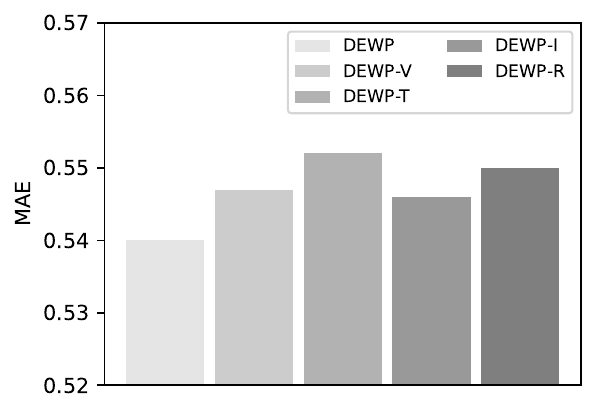} 
}  
\subfigure[\small lookback 24, R80721] {
  
\includegraphics[width=3.2cm]{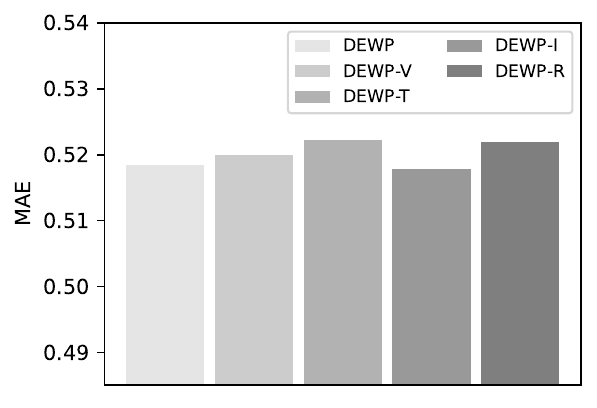}     
}
%\hspace{-2mm}
\subfigure[\small lookback 72, R80721] {
    
\includegraphics[width=3.2cm]{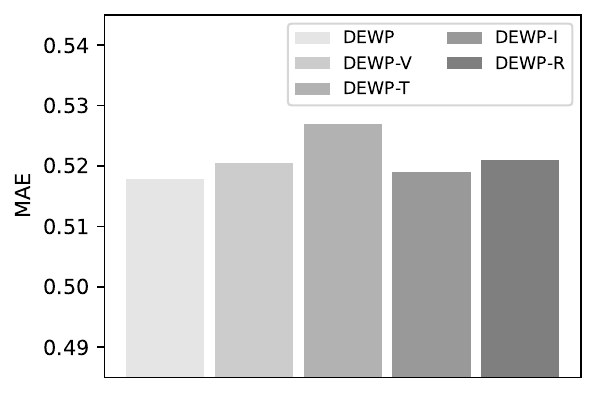}     
}
\caption{Ablation tests of Short-Term with different experimental setups on R80736 and R80721 datasets.}     
\label{fig:exp_ablation1}
\end{figure*}

\begin{figure*} \centering  
\subfigure[\small lookback 24, R80736] {
%\label{pa_exp1a}     
\includegraphics[width=3.2cm]{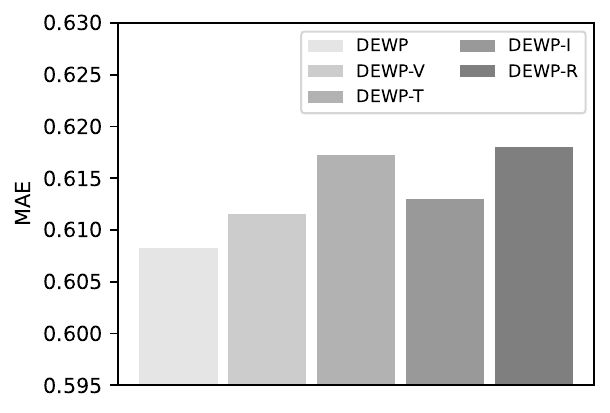} 
}  
%\hspace{-2mm}
\subfigure[\small lookback 72, R80736] {
   
\includegraphics[width=3.2cm]{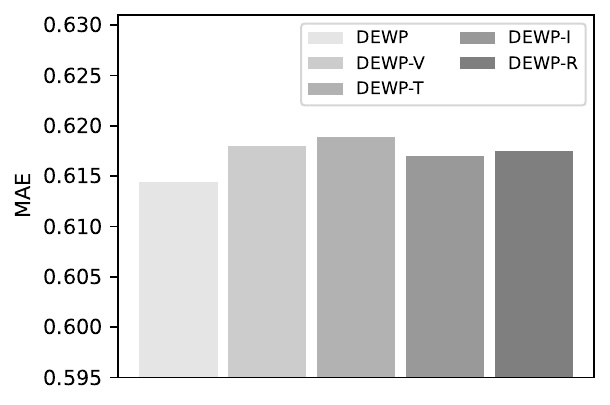} 
}  
\subfigure[\small lookback 24, R80721] {
  
\includegraphics[width=3.2cm]{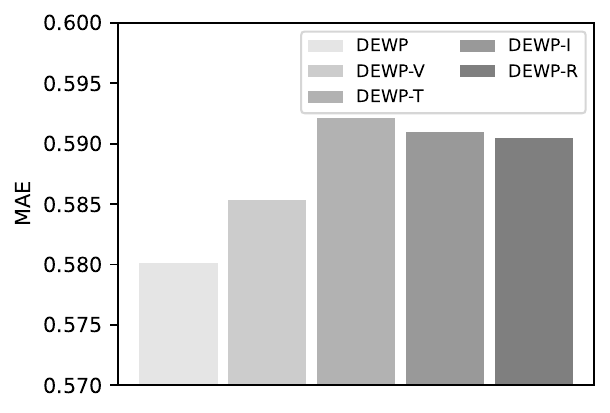}     
}
%\hspace{-2mm}
\subfigure[\small lookback 72, R80721] {
    
\includegraphics[width=3.2cm]{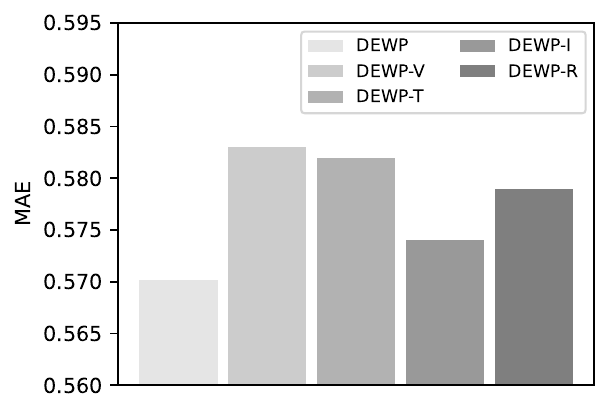}     
}
\caption{Ablation tests of Long-Term with different experimental setups on R80736 and R80721 datasets.}     
\label{fig:exp_ablation2}
\end{figure*}

\begin{figure*}
\centering 
\subfigure[\small R80736] {
\includegraphics[width=4cm]{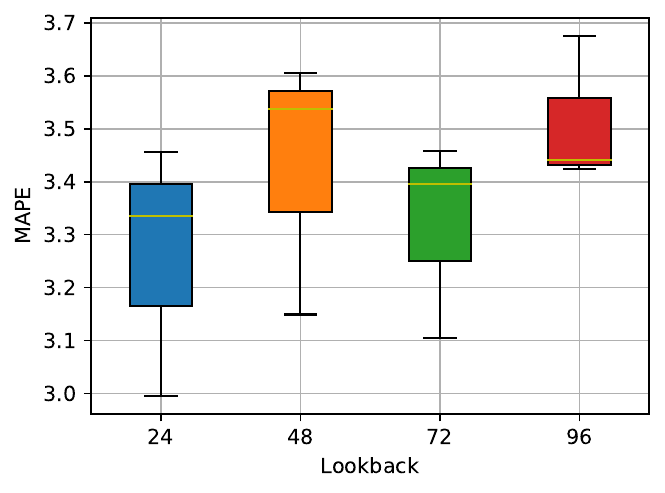} 
}  
\hspace{+6mm}
\subfigure[\small R80721] {
 \label{pa_exp1b}     
\includegraphics[width=4cm]{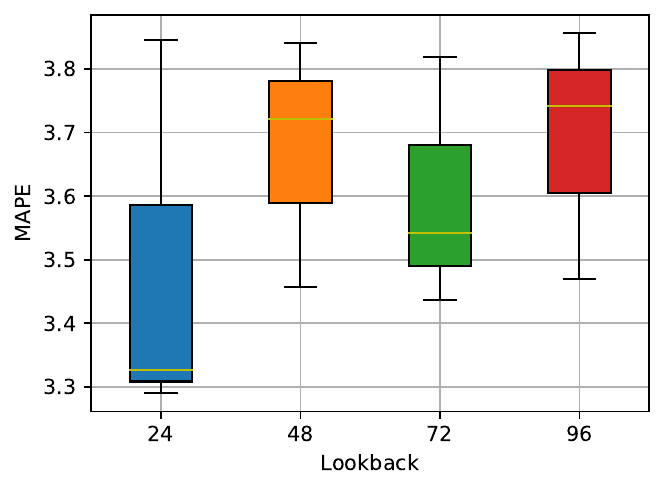} 
} 
\vspace{-3mm}
\caption{Parameter analysis towards lookback on R80736 and R80721 datasets. For the choice of lookback window length, we traverse from 24, 48, 72 to 96 to test the performances.}
\label{fig:pa1} 
\end{figure*}

\begin{figure*}
\centering  
\subfigure[\small R80736] {
\includegraphics[width=4cm]{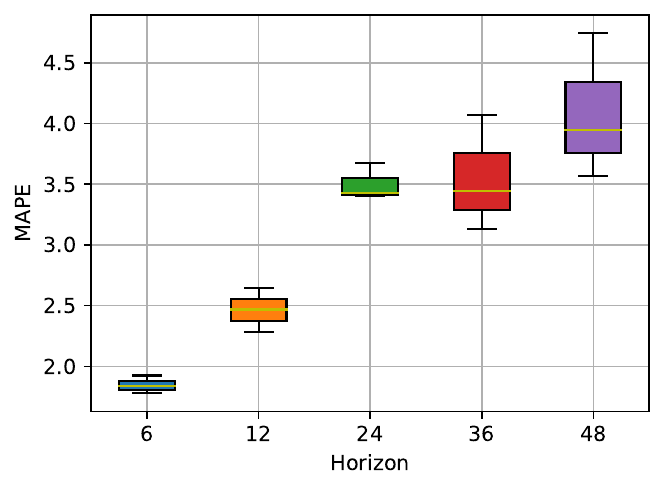} 
}  
\hspace{+6mm}
\subfigure[\small R80721] {
 \label{pa_exp1b}     
\includegraphics[width=4cm]{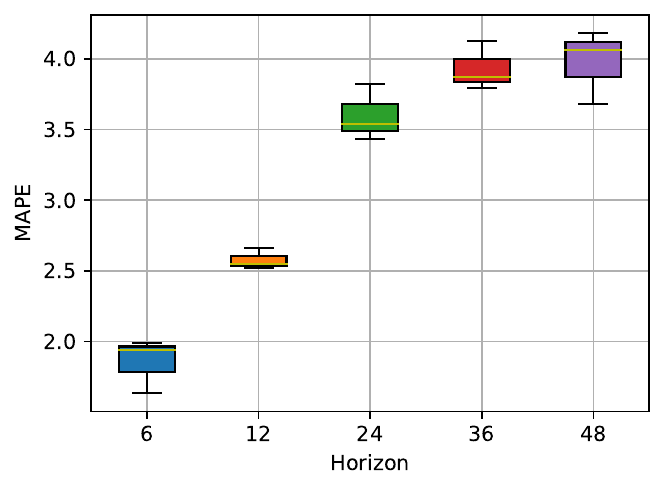} 
}
\vspace{-3mm}
\caption{Parameter analysis towards horizon on R80736 and R80721 datasets. For the choice of horizon window length, we traverse from 6 to 48 and test the performances.}  \label{fig:pa2} 
\end{figure*}

We evaluated performances of above variants in both \textsc{Short-Term} and \textsc{Long-Term} wind power forecasting on two datasets. 
Figure \ref{fig:exp_ablation1} and Figure \ref{fig:exp_ablation2}  demonstrate the performance comparisons of four variants and the original model in different setups. We easily observe these variants have lower performances than the original \textit{DEWP}; this demonstrates every component of the main model benefits and is important to final forecasting. Among all the variants, we found that \textit{DEWP-T} usually has a significant performance loss compared with the original model; this shows the forecasting largely depends on the time expansion modelling.  We noticed \textit{DEWP-R} can lead to a significant prediction loss, which demonstrates that the residue connections are indispensable to accurate forecasting.

\subsection{Parameter Analysis} \label{sec:parameters}
We noticed that different experimental settings and parameters largely influence the prediction results. We presented the analysis of parameter sensitivity towards the DEWP framework. 

\subsubsection{Lookback.} The length of lookback window determines the amount of historical information input to the model, which is an important factor for prediction.
If the lookback length is too large, more redundant observations will be taken into the model and damage the performance. If the lookback length is too small, there might be not enough observations to support accurate forecasting. In this regard, we studied the prediction performances versus. lookback length: Figure \ref{fig:pa1} shows the experimental results on two datasets when other parameters are fixed. The experimental result further verifies \textit{DEWP} has a strong ability to make full use of recent observations for accurate forecasting, even with a small lookback window (24).

\begin{figure*} \centering  
\subfigure[\small R80736] {
\includegraphics[width=4cm]{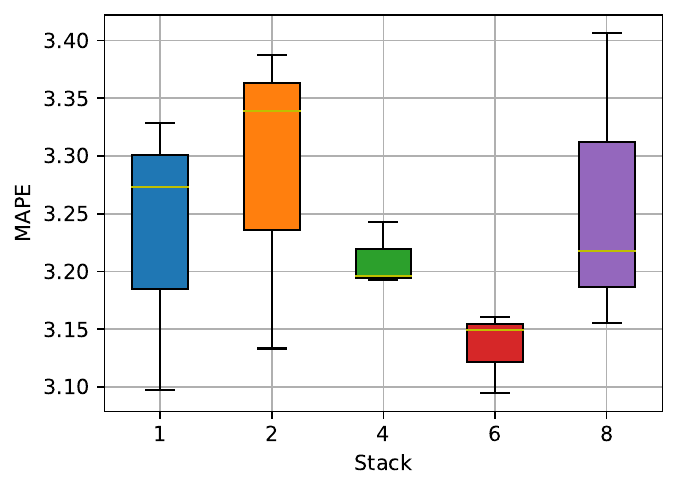} 
}  
\hspace{+6mm}
\subfigure[\small R80721] {
 \label{pa_exp1b}     
\includegraphics[width=4cm]{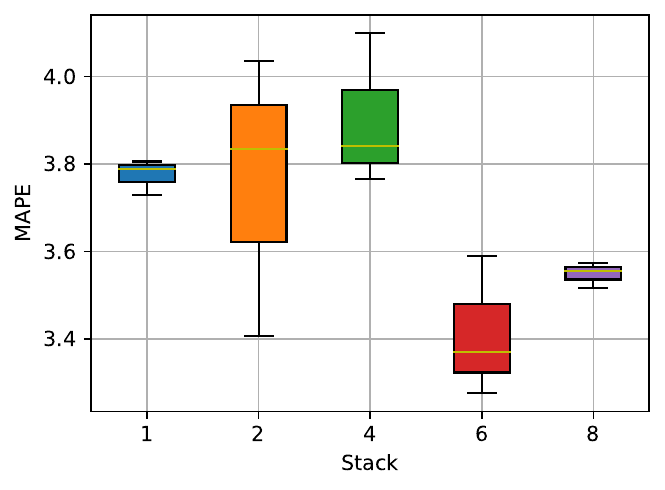} 
} 
\vspace{-3mm}
\caption{Parameter analysis towards stack on R80736 and R80721 datasets. For the choice of stack size, we traverse from 1 to 8 to test the performances.}  
\label{fig:pa3} 
\end{figure*}

\begin{figure*} \centering  
\subfigure[\small R80736] {
\includegraphics[width=4cm]{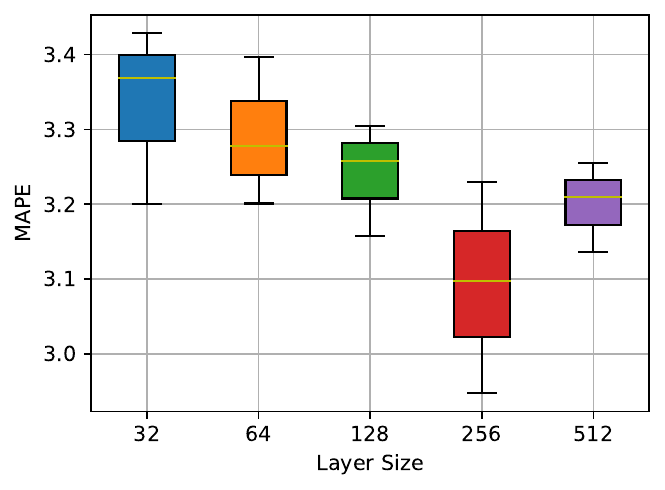} 
}  
\hspace{+6mm}
\subfigure[\small R80721] {
 \label{pa_exp1b}     
\includegraphics[width=4cm]{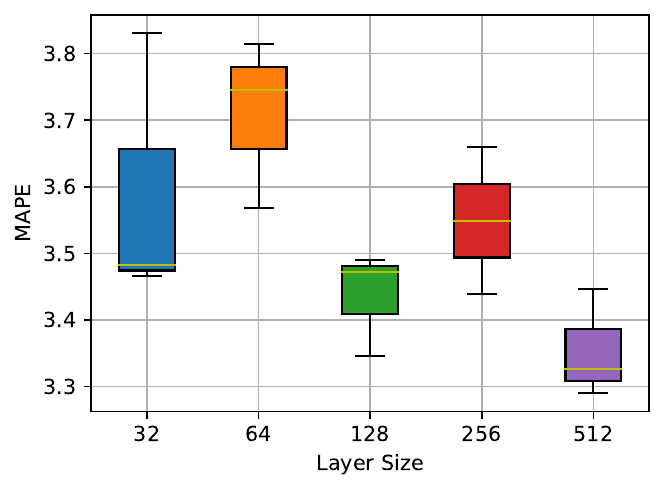} 
}
\vspace{-3mm}
\caption{Parameter analysis towards layer size on R80736 and R80721 datasets. For the choice of layer size, we traverse from 32 to 512 to test the performances.} \label{fig:pa4} 
\end{figure*}

\subsubsection{Horizon.} The length of forecast horizon determines the number of points to forecast. Fixing lookback equals to 72 and the rest parameters, we evaluated the performances of \textit{DEWP} with forecast horizon varying from 6 to 48. Figure \ref{fig:pa2} demonstrate the experimental results on two datasets.
%As a result, it is intuitive smaller horizon will decrease the prediction difficulty.
From the figures, we observed that larger forecast horizon makes larger prediction errors (MAPE). This is quite intuitive because larger horizon brings up more uncertainty to predict and increases prediction difficulty. As a result, long-term wind power forecasting leads to more errors than short-term forecasting. We used forecasting 24 hours ahead and 48 hours ahead as the main setting.

\subsubsection{Stack.} As introduced before, \textit{DEWP} is a stack-by-stack architecture, which has many stacked layers to make a generic and expressive model. In this regard, we studied the impacts of stacks on predictive performances by varying the number of stacks from 1 to 8. 
As Figure \ref{fig:pa3} shows, the performance improves when the number of stacks increases, which indicates that shallow layers have limitation in expressiveness, and deeper architectures can boost the forecasting to certain degree.

\subsubsection{Layer Size.} The layer size of \textit{DEWP} determines the number of neurons in hidden states of each block. We studied the size of inner layers of \textit{DEWP} by varying the size from 32 to 512, in order to compare their performances. Figure \ref{fig:pa4} shows the experimental results with five layer sizes on two datasets. We found that larger layer size usually brings a performance gain in prediction, but the gain is not so significant. Moreover, larger layer size leads to more memory consumption and training time; thus we need to balance efficiency and effectiveness for layer size setting.

\subsubsection{Training Duration.} We have collected the training time of our DEWP model under different experimental settings, including short-term window power forecasting and long-term wind power forecasting. Table \ref{table:tt1} and Table \ref{table:tt2} show the overall training time and training speed of short-term and long-term wind power forecasting. From the tables, we notice that with the increase of length of lookback windows, the training time is slightly enlarged, which signifies the larger input would cause more computation resource.
Moreover, we notice that with the increase of horizon length, long-term forecasting requires more training time and the training speed is also slightly slower than short-term forecasting, which is also acceptable.

\begin{table}[h!]
\centering
\caption{Training time and training speed of \textsc{Short-Term} wind power forecasting. The training speed is by calculating the training time of each iteration.} \label{table:tt1}
 \begin{tabular}{c c c } 
 \hline
 Lookback&Training Time & Training Speed\\
 \hline
 24 & 169.1245s&0.0160s/iter \\
 72 & 187.9539s&0.0177s/iter \\
 \hline
 \end{tabular}
\end{table}

\begin{table}[h!]
\centering
\caption{Training time and training speed of \textsc{Long-Term} wind power forecasting. The training speed is by calculating the training time of each iteration.} \label{table:tt2}
 \begin{tabular}{c c c } 
 \hline
 Lookback&Training Time & Training Speed\\
 \hline
 48 & 190.4898s&0.0180s/iter \\
 72 & 205.8578s&0.0194s/iter \\
 \hline
 \end{tabular}
\end{table}

\section{Experimental Discussions}

\subsection{Time Complexity Analysis}
DEWP is composed of several stacked layers, each of which is composed of one Variable Expansion Block, one Time Expansion Block, and one Inference Block. For the Variable Expansion Block, we adopt four 1-dimensional convolution layers. The complexity of convolution layers is $\mathcal{O}(k)$ given the kernel size of $k$. As a result, operating $n$ times convolution for a given time series infers to $\mathcal{O}(nk)$. Given the length of time series windows of length $L$, we have $n =[L/k]$ and we can equally have the complexity of variation expansion block is $\mathcal{O}([L/k] \times k) \approx \mathcal{O}(L)$.
For the time expansion block, we adopt the fully-connected layers for both generating coefficients and calculating Fourier series. Accordingly, the complexity of time expansion block is approximately equal to $\mathcal{O}(L)$ which is the complexity of linear layers.
In addition, we adopt multi-head self-attention mechanisms in the inference block and the complexity of self-attention computation comes to $\mathcal{O}(L^2)$ \cite{vaswani2017attention}. In summary, the complexity of DEWP is $\mathcal{O}(L)+ \mathcal{O}(L) + \mathcal{O}(L^2) \approx \mathcal{O}(L^2)$.

\subsection{Deployment Plan}
DEWP is developed to provide robust and accurate wind power forecasting for industrial partners. We present Algorithm \ref{algorithm1} that describes the details of training procedure of DEWP.
Given the evidence provided, our algorithm will be deployed on the backend to automatically predict the wind power given the natural observations. 

Under a new wind power forecasting situation, to begin with, all the collected historical data can be used to initially train the DEWP base model.
Then, DEWP can be re-trained frequently, such as weekly or monthly, using new upcoming data to update the base model.
During the real-world deployment, several DEWP models with different setup (different seeds, different lookback windows, different horizons) can be trained in order to report a robust ensembled results. After the backend results are acquired, they can be shown up in the frontend for further analysis.

\subsection{Managerial Implications}
%For different situations, 
Generally, the electricity market is build upon two mechanisms.
The first one is Day Ahead market, where the bulk energy necessary to cover the demand for the next day is traded on the market. The accurate wind power forecasting permits the settlement of electricity price ahead of time during the bidding and the auction process.
The second mechanism is ancillary service market, where the gap between planned production and actual load is traded (due to the power plant failure or due to intermittence of wind power generation). The accurate wind power forecasting can support a stable operation of the power grid by arranging power production in the ancillary service market. It is very important for consumers and suppliers because they can imply the future electricity price ahead of time and make strategies accordingly.

\begin{algorithm}
\caption{DEWP training procedure.}\label{algorithm1}
\KwIn{Multi-variate natural observations $\mathbf{X} \in \mathbb{R}^{d*l}$, Actual generated wind power data $\mathbf{y} \in \mathbb{R}^l$, where $d$ is the number of variables and $l$ is length of training series.}
\KwOut{Predicted wind power generation $\hat{\mathbf{y}}$}
% $r\leftarrow t$\;
% $\Delta B^{\ast}\leftarrow -\infty$\;
\While{not converged}
{ 
Sample a lookback window data of length $L$ for $b$ times to make a batch $\mathbf{X}^B_{t-L:t} \in \mathbb{R}^{b*d*L}$ \; 
Sample the corresponding label batch $\mathbf{y}^B_{t:t+H} \in \mathbb{R}^{b*H}$ with forecast horizon as $H$ \;
Initialize $\mathbf{X}^{(0)}_{t-L:t} = \mathbf{X}^B_{t-L:t}$, Initialize $ \hat{\mathbf{y}}_{t:t+H} = 0$ \;
\For{$\ell\gets1$ \KwTo $M$}{
    backcast $\widehat{\mathbf{X}}_b^{(\ell)}$, forecast $\widehat{\mathbf{X}}_f^{(\ell)} = StackProcess(\mathbf{X}^{(\ell)}_{t-L:t})$ \;
    $\hat{\mathbf{y}}^{(\ell)}_{f} = InferenceBlock(\widehat{\mathbf{X}}_f^{(\ell)})$ \;
    $\mathbf{X}^{(\ell+1)}_{t-L:t} = \mathbf{X}^{(\ell)}_{t-L:t} - \widehat{\mathbf{X}}_b^{(\ell)}$ \;
    $\hat{\mathbf{y}}_{t:t+H} = \hat{\mathbf{y}}_{t:t+H} +  \hat{\mathbf{y}}^{(\ell)}_{f}$ \;
    }
Calculate loss $\mathcal{L} = MeanSquareError(\hat{\mathbf{y}}_{t:t+H}, \mathbf{y}^B_{t:t+H})$ \;
Update parameters of stacks and inference blocks \;
}
\end{algorithm}

\subsection{Limitations and Future Work}
In this section, we discuss the limitations of DEWP and also some future thoughts.
The main limitations lie in two aspects:
First, DEWP is a deep-learning based model, which cannot take real-time data as input. Thus, we need to retrain the models after a period of time, in order to keep the data updated with the upcoming data. Second, the inference block applies self attention mechanisms, which requires enough computational resources in a $O(N^2)$ complexity.
The two limitations naturally inspire us to improve DEWP from these perspectives: (i) improve DEWP with some real-time algorithms; (ii) improve the computation efficiency with other mechanisms (e.g., more efficient attention) in inference blocks.

\section{Related Work}

\subsection{Time Series Forecasting}
Time series forecasting is a classic problem  that has been extensively studied for decades. 
At the beginning, researchers developed simple statistical modeling methods, such as  exponentially weighted moving averages~\cite{holt2004forecasting}, auto-regressive moving averages (ARMA)~\cite{whittle1963prediction}.
However, these statistical approaches only considered simple linear dependencies of future signals on historical observations. 
Later, with the great successes of deep neural networks (DNNs), many DNN-based methods have been proposed. 
One kind of methods is to use recurrent neural networks (RNN) for time series forecasting \cite{salinas2020deepar}. As RNN's variants, Long Short-Term Memory (LSTM) \cite{gers2002applying} or and Gated Recurrent Unit (GRU) \cite{chung2014empirical} become more popular for series data.
Due to the effectiveness of the self-attention mechanisms, Transformer \cite{vaswani2017attention} and its improved version like Informer \cite{zhou2021informer} or Autoformer \cite{xu2021autoformer} have taken the place of RNN models many sequence modeling tasks~\cite{yi2023survey}.
Other researchers also tried to apply convolution networks \cite{bai2018empirical,ye2023web}, graph neural networks~\cite{yi2023fouriergnn,yi2023survey} or pure fully-connected networks \cite{oreshkin2019n,fan2022depts,fan2023dish,yi2023frequency}, showing much effectiveness in time series forecasting.

\subsection{Wind Power Forecasting}
As a special kind of time series task, Wind Power Forecasting (WPF) has received significant attention due to its practical value in industry and society.
Based on the time-scale, WPF approaches can be grouped into immediate-short-term (8 hours-ahead) forecasting, short-term (day-ahead) forecasting, long-term (multiple-days-ahead) forecasting \cite{wang2011review}.
There exist different kinds of approaches for WPF \cite{maldonado2021wind}: (i) \textit{Physical} approaches use physical characterisation to model wind turbines/farms based on the numerical weather prediction (NWP) data \cite{lange2008new}; 
(ii) \textit{Statistical} methods construct linear relationships between NWPs data and the generated power and use autoregressive processes to  model such relationships: for example, \cite{yuan2017wind} proposes to use ARIMA (autoregressive integrated moving average) together with least-
squares support-vector machine (LS-SVM) to model the linear component of WP time series; \cite{xie2018nonparametric} proposes a non-parametric statistical model based on Markov state transition process;
(iii) \textit{Deep Learning}-based methods apply deep neural networks into the wind power forecasting: Some methods are based on recurrent neural network (RNN) such as LSTM-EFG \cite{yu2019lstm}, Gated Recurrent Neural Network \cite{kisvari2021wind}, HBO-LSTM \cite{ewees2022hbo};
Some methods are based on convolution neural network (CNN) such as temporal convolution network \cite{gan2021temporal}, conformalized temporal convolutional quantile regression networks~\cite{hu2022conformalized} and  an improved residual-based convolutional neural network \cite{yildiz2021improved};
Recently, the development of attention mechanisms has fostered many attention-based methods such as dual-stage self-attention network~\cite{tian2022developing}, temporal fusion transformers for wind prediction~\cite{wu2022interpretable}, etc.
In addition, some methods are hybrid methods composed of different networks, such as decomposition network and convolution network \cite{li2022multi}, recurrent network and attention network \cite{niu2022point,han2023machine}. Though many mechanisms have been used, their sophisticated architectures would potentially bring expensive computation consumption. And they don't construct the multi-view dependencies of future wind power on the historical observations which hinders the prediction performances.

%However, they always simply adopt the vanilla model and don't construct complicated dependencies of future wind power on historical observations which hinders the prediction performances.

\section{Conclusion}
In this paper, we presented DEWP, a novel deep expansion learning framework for wind power forecasting. 
Our model is built upon several tailored-designed blocks:
variable expansion blocks captured dependencies among variables and expanded input for feature extraction;
time expansion blocks captured dependencies in temporal patterns and expanded representations to backcast and forecast;
inference blocks carefully mapped expanded forecasts to the target wind power values.
All these blocks were coupled with doubly residue learning for better expansions.
Extensive experiments were conducted on real-world wind power datasets to prove the effectiveness of our method.
As a successful neural forecasting tool, our DEWP model will be deployed on a wind power data mining platform, in order to provide energy generation prediction towards better plant operation and management.
%Currently, we are exploring better solutions to improve wind power forecasting for a series of turbines, which is more challenging since we need to 

% \input{7_appendix}

\begin{acks}
This work was partially done at the University of Central Florida. This work was partially supported by the Natural Science Foundation of China (No. 61836013), National Natural Science
Foundation of China (Grant No.92370204), Guangzhou-HKUST(GZ) Joint Funding 
Program (Grant No.2023A03J0008), Education Bureau of Guangzhou Municipality, and Guangdong Science and Technology Department.

% This research was partially supported by the National Science Foundation (NSF) via the grant numbers: 2040950, 2006889, 2045567.
\end{acks}

\bibliographystyle{ACM-Reference-Format}
\bibliography{sample-base}


\begin{thebibliography}{53}

%%% the Web address.

\ifx \showCODEN    \undefined \def \showCODEN     #1{\unskip}     \fi
\ifx \showDOI      \undefined \def \showDOI       #1{#1}\fi
\ifx \showISBNx    \undefined \def \showISBNx     #1{\unskip}     \fi
\ifx \showISBNxiii \undefined \def \showISBNxiii  #1{\unskip}     \fi
\ifx \showISSN     \undefined \def \showISSN      #1{\unskip}     \fi
\ifx \showLCCN     \undefined \def \showLCCN      #1{\unskip}     \fi
\ifx \shownote     \undefined \def \shownote      #1{#1}          \fi
\ifx \showarticletitle \undefined \def \showarticletitle #1{#1}   \fi
\ifx \showURL      \undefined \def \showURL       {\relax}        \fi
% invisible to TeX
\providecommand\bibfield[2]{#2}
\providecommand\bibinfo[2]{#2}
\providecommand\natexlab[1]{#1}
\providecommand\showeprint[2][]{arXiv:#2}

\bibitem[Bai et~al\mbox{.}(2018)]%
        {bai2018empirical}
\bibfield{author}{\bibinfo{person}{Shaojie Bai}, \bibinfo{person}{J~Zico
  Kolter}, {and} \bibinfo{person}{Vladlen Koltun}.}
  \bibinfo{year}{2018}\natexlab{}.
\newblock \showarticletitle{An empirical evaluation of generic convolutional
  and recurrent networks for sequence modeling}.
\newblock  (\bibinfo{year}{2018}), \bibinfo{pages}{1--12}.
\newblock


\bibitem[Bilal et~al\mbox{.}(2018)]%
        {bilal2018wind}
\bibfield{author}{\bibinfo{person}{Boudy Bilal}, \bibinfo{person}{Mamoudou
  Ndongo}, \bibinfo{person}{Kondo~H Adjallah}, \bibinfo{person}{Alexandre
  Sava}, \bibinfo{person}{Cheikh~MF K{\'e}b{\'e}},
  \bibinfo{person}{Pape~Alioune Ndiaye}, {and} \bibinfo{person}{Vincent
  Sambou}.} \bibinfo{year}{2018}\natexlab{}.
\newblock \showarticletitle{Wind turbine power output prediction model design
  based on artificial neural networks and climatic spatiotemporal data}. In
  \bibinfo{booktitle}{\emph{2018 IEEE International Conference on Industrial
  Technology (ICIT)}}. IEEE, \bibinfo{pages}{1085--1092}.
\newblock


\bibitem[Borovykh et~al\mbox{.}(2017)]%
        {borovykh2017conditional}
\bibfield{author}{\bibinfo{person}{Anastasia Borovykh}, \bibinfo{person}{Sander
  Bohte}, {and} \bibinfo{person}{Cornelis~W Oosterlee}.}
  \bibinfo{year}{2017}\natexlab{}.
\newblock \showarticletitle{Conditional time series forecasting with
  convolutional neural networks}.
\newblock \bibinfo{journal}{\emph{Journal of Computational Finance}}
  \bibinfo{number}{4} (\bibinfo{year}{2017}), \bibinfo{pages}{1--12}.
\newblock


\bibitem[Chung et~al\mbox{.}(2014)]%
        {chung2014empirical}
\bibfield{author}{\bibinfo{person}{Junyoung Chung}, \bibinfo{person}{Caglar
  Gulcehre}, \bibinfo{person}{KyungHyun Cho}, {and} \bibinfo{person}{Yoshua
  Bengio}.} \bibinfo{year}{2014}\natexlab{}.
\newblock \showarticletitle{Empirical evaluation of gated recurrent neural
  networks on sequence modeling}.
\newblock  (\bibinfo{year}{2014}), \bibinfo{pages}{1--12}.
\newblock


\bibitem[El-Ahmar et~al\mbox{.}(2017)]%
        {el2017evaluation}
\bibfield{author}{\bibinfo{person}{MH El-Ahmar},
  \bibinfo{person}{Abou-Hashema~M El-Sayed}, {and} \bibinfo{person}{AM
  Hemeida}.} \bibinfo{year}{2017}\natexlab{}.
\newblock \showarticletitle{Evaluation of factors affecting wind turbine output
  power}. In \bibinfo{booktitle}{\emph{2017 Nineteenth International Middle
  East Power Systems Conference (MEPCON)}}. IEEE, \bibinfo{pages}{1471--1476}.
\newblock


\bibitem[Ewees et~al\mbox{.}(2022)]%
        {ewees2022hbo}
\bibfield{author}{\bibinfo{person}{Ahmed~A Ewees}, \bibinfo{person}{Mohammed~AA
  Al-qaness}, \bibinfo{person}{Laith Abualigah}, {and} \bibinfo{person}{Mohamed
  Abd~Elaziz}.} \bibinfo{year}{2022}\natexlab{}.
\newblock \showarticletitle{HBO-LSTM: Optimized long short term memory with
  heap-based optimizer for wind power forecasting}.
\newblock \bibinfo{journal}{\emph{Energy Conversion and Management}}
  \bibinfo{volume}{268} (\bibinfo{year}{2022}), \bibinfo{pages}{116022}.
\newblock
\showISSN{0196-8904}


\bibitem[Fan et~al\mbox{.}(2021)]%
        {fan2021interactive}
\bibfield{author}{\bibinfo{person}{Wei Fan}, \bibinfo{person}{Kunpeng Liu},
  \bibinfo{person}{Hao Liu}, \bibinfo{person}{Yong Ge}, \bibinfo{person}{Hui
  Xiong}, {and} \bibinfo{person}{Yanjie Fu}.} \bibinfo{year}{2021}\natexlab{}.
\newblock \showarticletitle{Interactive reinforcement learning for feature
  selection with decision tree in the loop}.
\newblock \bibinfo{journal}{\emph{IEEE Transactions on Knowledge and Data
  Engineering}} (\bibinfo{year}{2021}).
\newblock


\bibitem[Fan et~al\mbox{.}(2020)]%
        {fan2020autofs}
\bibfield{author}{\bibinfo{person}{Wei Fan}, \bibinfo{person}{Kunpeng Liu},
  \bibinfo{person}{Hao Liu}, \bibinfo{person}{Pengyang Wang},
  \bibinfo{person}{Yong Ge}, {and} \bibinfo{person}{Yanjie Fu}.}
  \bibinfo{year}{2020}\natexlab{}.
\newblock \showarticletitle{Autofs: Automated feature selection via
  diversity-aware interactive reinforcement learning}. In
  \bibinfo{booktitle}{\emph{2020 IEEE International Conference on Data Mining
  (ICDM)}}. IEEE, \bibinfo{pages}{1008--1013}.
\newblock


\bibitem[Fan et~al\mbox{.}(2023)]%
        {fan2023dish}
\bibfield{author}{\bibinfo{person}{Wei Fan}, \bibinfo{person}{Pengyang Wang},
  \bibinfo{person}{Dongkun Wang}, \bibinfo{person}{Dongjie Wang},
  \bibinfo{person}{Yuanchun Zhou}, {and} \bibinfo{person}{Yanjie Fu}.}
  \bibinfo{year}{2023}\natexlab{}.
\newblock \showarticletitle{Dish-TS: a general paradigm for alleviating
  distribution shift in time series forecasting}. In
  \bibinfo{booktitle}{\emph{Proceedings of the AAAI Conference on Artificial
  Intelligence}}, Vol.~\bibinfo{volume}{37}. \bibinfo{pages}{7522--7529}.
\newblock


\bibitem[Fan et~al\mbox{.}(2022)]%
        {fan2022depts}
\bibfield{author}{\bibinfo{person}{Wei Fan}, \bibinfo{person}{Shun Zheng},
  \bibinfo{person}{Xiaohan Yi}, \bibinfo{person}{Wei Cao},
  \bibinfo{person}{Yanjie Fu}, \bibinfo{person}{Jiang Bian}, {and}
  \bibinfo{person}{Tie-Yan Liu}.} \bibinfo{year}{2022}\natexlab{}.
\newblock \showarticletitle{{DEPTS}: Deep Expansion Learning for Periodic Time
  Series Forecasting}. In \bibinfo{booktitle}{\emph{International Conference on
  Learning Representations}}.
\newblock


\bibitem[Gan et~al\mbox{.}(2021)]%
        {gan2021temporal}
\bibfield{author}{\bibinfo{person}{Zhenhao Gan}, \bibinfo{person}{Chaoshun Li},
  \bibinfo{person}{Jianzhong Zhou}, {and} \bibinfo{person}{Geng Tang}.}
  \bibinfo{year}{2021}\natexlab{}.
\newblock \showarticletitle{Temporal convolutional networks interval prediction
  model for wind speed forecasting}.
\newblock \bibinfo{journal}{\emph{Electric Power Systems Research}}
  \bibinfo{volume}{191} (\bibinfo{year}{2021}), \bibinfo{pages}{106865}.
\newblock
\showISSN{0378-7796}


\bibitem[Gers et~al\mbox{.}(2002)]%
        {gers2002applying}
\bibfield{author}{\bibinfo{person}{Felix~A Gers}, \bibinfo{person}{Douglas
  Eck}, {and} \bibinfo{person}{J{\"u}rgen Schmidhuber}.}
  \bibinfo{year}{2002}\natexlab{}.
\newblock \showarticletitle{Applying LSTM to time series predictable through
  time-window approaches}.
\newblock In \bibinfo{booktitle}{\emph{Neural Nets WIRN Vietri-01}}.
  \bibinfo{publisher}{Springer}, \bibinfo{pages}{193--200}.
\newblock


\bibitem[Giebel et~al\mbox{.}(2006)]%
        {giebel2006shortterm}
\bibfield{author}{\bibinfo{person}{G Giebel}, \bibinfo{person}{J Badger},
  \bibinfo{person}{I~Mart{\'\i} Perez}, \bibinfo{person}{P Louka},
  \bibinfo{person}{G Kallos}, \bibinfo{person}{AM Palomares},
  \bibinfo{person}{C Lac}, {and} \bibinfo{person}{G Descombes}.}
  \bibinfo{year}{2006}\natexlab{}.
\newblock \showarticletitle{Shortterm forecasting using advanced physical
  modelling-the results of the anemos project}. In
  \bibinfo{booktitle}{\emph{Proceedings of the European Wind Energy
  Conference}}.
\newblock


\bibitem[Giebel et~al\mbox{.}(2011)]%
        {giebel2011state}
\bibfield{author}{\bibinfo{person}{Gregor Giebel}, \bibinfo{person}{Caroline
  Draxl}, \bibinfo{person}{Richard Brownsword}, \bibinfo{person}{Georges
  Kariniotakis}, {and} \bibinfo{person}{Michael Denhard}.}
  \bibinfo{year}{2011}\natexlab{}.
\newblock \showarticletitle{The state-of-the-art in short-term prediction of
  wind power. A literature overview}.
\newblock   \bibinfo{volume}{0} (\bibinfo{year}{2011}),
  \bibinfo{pages}{1--111}.
\newblock


\bibitem[Gielen et~al\mbox{.}(2019)]%
        {gielen2019role}
\bibfield{author}{\bibinfo{person}{Dolf Gielen}, \bibinfo{person}{Francisco
  Boshell}, \bibinfo{person}{Deger Saygin}, \bibinfo{person}{Morgan~D
  Bazilian}, \bibinfo{person}{Nicholas Wagner}, {and} \bibinfo{person}{Ricardo
  Gorini}.} \bibinfo{year}{2019}\natexlab{}.
\newblock \showarticletitle{The role of renewable energy in the global energy
  transformation}.
\newblock \bibinfo{journal}{\emph{Energy Strategy Reviews}}
  \bibinfo{volume}{24} (\bibinfo{year}{2019}), \bibinfo{pages}{38--50}.
\newblock
\showISSN{2211-467X}


\bibitem[Glorot et~al\mbox{.}(2011)]%
        {glorot2011deep}
\bibfield{author}{\bibinfo{person}{Xavier Glorot}, \bibinfo{person}{Antoine
  Bordes}, {and} \bibinfo{person}{Yoshua Bengio}.}
  \bibinfo{year}{2011}\natexlab{}.
\newblock \showarticletitle{Deep sparse rectifier neural networks}. In
  \bibinfo{booktitle}{\emph{Proceedings of the fourteenth international
  conference on artificial intelligence and statistics}}. JMLR Workshop and
  Conference Proceedings, \bibinfo{pages}{315--323}.
\newblock


\bibitem[Han et~al\mbox{.}(2023)]%
        {han2023machine}
\bibfield{author}{\bibinfo{person}{Jindong Han}, \bibinfo{person}{Weijia
  Zhang}, \bibinfo{person}{Hao Liu}, {and} \bibinfo{person}{Hui Xiong}.}
  \bibinfo{year}{2023}\natexlab{}.
\newblock \showarticletitle{Machine Learning for Urban Air Quality Analytics: A
  Survey}.
\newblock \bibinfo{journal}{\emph{arXiv preprint arXiv:2310.09620}}
  (\bibinfo{year}{2023}).
\newblock


\bibitem[Hanifi et~al\mbox{.}(2020)]%
        {hanifi2020critical}
\bibfield{author}{\bibinfo{person}{Shahram Hanifi}, \bibinfo{person}{Xiaolei
  Liu}, \bibinfo{person}{Zi Lin}, {and} \bibinfo{person}{Saeid Lotfian}.}
  \bibinfo{year}{2020}\natexlab{}.
\newblock \showarticletitle{A critical review of wind power forecasting
  methods—past, present and future}.
\newblock \bibinfo{journal}{\emph{Energies}} \bibinfo{volume}{13},
  \bibinfo{number}{15} (\bibinfo{year}{2020}), \bibinfo{pages}{3764}.
\newblock


\bibitem[He et~al\mbox{.}(2016)]%
        {he2016deep}
\bibfield{author}{\bibinfo{person}{Kaiming He}, \bibinfo{person}{Xiangyu
  Zhang}, \bibinfo{person}{Shaoqing Ren}, {and} \bibinfo{person}{Jian Sun}.}
  \bibinfo{year}{2016}\natexlab{}.
\newblock \showarticletitle{Deep residual learning for image recognition}. In
  \bibinfo{booktitle}{\emph{Proceedings of the IEEE conference on computer
  vision and pattern recognition}}. \bibinfo{pages}{770--778}.
\newblock


\bibitem[Hemami(2012)]%
        {hemami2012wind}
\bibfield{author}{\bibinfo{person}{Ahmad Hemami}.}
  \bibinfo{year}{2012}\natexlab{}.
\newblock \bibinfo{booktitle}{\emph{Wind turbine technology}}.
\newblock \bibinfo{publisher}{Cengage Learning}.
\newblock


\bibitem[Holt(2004)]%
        {holt2004forecasting}
\bibfield{author}{\bibinfo{person}{Charles~C Holt}.}
  \bibinfo{year}{2004}\natexlab{}.
\newblock \showarticletitle{Forecasting seasonals and trends by exponentially
  weighted moving averages}.
\newblock \bibinfo{journal}{\emph{International journal of forecasting}}
  \bibinfo{volume}{20}, \bibinfo{number}{1} (\bibinfo{year}{2004}),
  \bibinfo{pages}{5--10}.
\newblock


\bibitem[Hong and Rioflorido(2019)]%
        {hong2019hybrid}
\bibfield{author}{\bibinfo{person}{Ying-Yi Hong} {and}
  \bibinfo{person}{Christian Lian Paulo~P Rioflorido}.}
  \bibinfo{year}{2019}\natexlab{}.
\newblock \showarticletitle{A hybrid deep learning-based neural network for
  24-h ahead wind power forecasting}.
\newblock \bibinfo{journal}{\emph{Applied Energy}}  \bibinfo{volume}{250}
  (\bibinfo{year}{2019}), \bibinfo{pages}{530--539}.
\newblock


\bibitem[Hu et~al\mbox{.}(2022)]%
        {hu2022conformalized}
\bibfield{author}{\bibinfo{person}{Jianming Hu}, \bibinfo{person}{Qingxi Luo},
  \bibinfo{person}{Jingwei Tang}, \bibinfo{person}{Jiani Heng}, {and}
  \bibinfo{person}{Yuwen Deng}.} \bibinfo{year}{2022}\natexlab{}.
\newblock \showarticletitle{Conformalized temporal convolutional quantile
  regression networks for wind power interval forecasting}.
\newblock \bibinfo{journal}{\emph{Energy}}  \bibinfo{volume}{248}
  (\bibinfo{year}{2022}), \bibinfo{pages}{123497}.
\newblock


\bibitem[Kisvari et~al\mbox{.}(2021)]%
        {kisvari2021wind}
\bibfield{author}{\bibinfo{person}{Adam Kisvari}, \bibinfo{person}{Zi Lin},
  {and} \bibinfo{person}{Xiaolei Liu}.} \bibinfo{year}{2021}\natexlab{}.
\newblock \showarticletitle{Wind power forecasting--A data-driven method along
  with gated recurrent neural network}.
\newblock \bibinfo{journal}{\emph{Renewable Energy}}  \bibinfo{volume}{163}
  (\bibinfo{year}{2021}), \bibinfo{pages}{1895--1909}.
\newblock


\bibitem[Lai et~al\mbox{.}(2018)]%
        {lai2018modeling}
\bibfield{author}{\bibinfo{person}{Guokun Lai}, \bibinfo{person}{Wei-Cheng
  Chang}, \bibinfo{person}{Yiming Yang}, {and} \bibinfo{person}{Hanxiao Liu}.}
  \bibinfo{year}{2018}\natexlab{}.
\newblock \showarticletitle{Modeling long-and short-term temporal patterns with
  deep neural networks}. In \bibinfo{booktitle}{\emph{The 41st International
  ACM SIGIR Conference on Research \& Development in Information Retrieval}}.
  \bibinfo{pages}{95--104}.
\newblock


\bibitem[Lange and Focken(2008)]%
        {lange2008new}
\bibfield{author}{\bibinfo{person}{Matthias Lange} {and}
  \bibinfo{person}{Ulrich Focken}.} \bibinfo{year}{2008}\natexlab{}.
\newblock \showarticletitle{New developments in wind energy forecasting}. In
  \bibinfo{booktitle}{\emph{2008 IEEE power and energy society general
  meeting-conversion and delivery of electrical energy in the 21st century}}.
  IEEE, \bibinfo{pages}{1--8}.
\newblock


\bibitem[LeCun et~al\mbox{.}(1995)]%
        {lecun1995convolutional}
\bibfield{author}{\bibinfo{person}{Yann LeCun}, \bibinfo{person}{Yoshua
  Bengio}, {et~al\mbox{.}}} \bibinfo{year}{1995}\natexlab{}.
\newblock \showarticletitle{Convolutional networks for images, speech, and time
  series}.
\newblock \bibinfo{journal}{\emph{The handbook of brain theory and neural
  networks}} \bibinfo{volume}{3361}, \bibinfo{number}{10}
  (\bibinfo{year}{1995}), \bibinfo{pages}{1995}.
\newblock


\bibitem[Li et~al\mbox{.}(2019)]%
        {li2019short}
\bibfield{author}{\bibinfo{person}{Chaoshun Li}, \bibinfo{person}{Geng Tang},
  \bibinfo{person}{Xiaoming Xue}, \bibinfo{person}{Adnan Saeed}, {and}
  \bibinfo{person}{Xin Hu}.} \bibinfo{year}{2019}\natexlab{}.
\newblock \showarticletitle{Short-term wind speed interval prediction based on
  ensemble GRU model}.
\newblock \bibinfo{journal}{\emph{IEEE transactions on sustainable energy}}
  \bibinfo{volume}{11}, \bibinfo{number}{3} (\bibinfo{year}{2019}),
  \bibinfo{pages}{1370--1380}.
\newblock


\bibitem[Li et~al\mbox{.}(2022)]%
        {li2022multi}
\bibfield{author}{\bibinfo{person}{Dan Li}, \bibinfo{person}{Fuxin Jiang},
  \bibinfo{person}{Min Chen}, {and} \bibinfo{person}{Tao Qian}.}
  \bibinfo{year}{2022}\natexlab{}.
\newblock \showarticletitle{Multi-step-ahead wind speed forecasting based on a
  hybrid decomposition method and temporal convolutional networks}.
\newblock \bibinfo{journal}{\emph{Energy}}  \bibinfo{volume}{238}
  (\bibinfo{year}{2022}), \bibinfo{pages}{121981}.
\newblock


\bibitem[Mahoney et~al\mbox{.}(2012)]%
        {mahoney2012wind}
\bibfield{author}{\bibinfo{person}{William~P Mahoney}, \bibinfo{person}{Keith
  Parks}, \bibinfo{person}{Gerry Wiener}, \bibinfo{person}{Yubao Liu},
  \bibinfo{person}{William~L Myers}, \bibinfo{person}{Juanzhen Sun},
  \bibinfo{person}{Luca Delle~Monache}, \bibinfo{person}{Thomas Hopson},
  \bibinfo{person}{David Johnson}, {and} \bibinfo{person}{Sue~Ellen Haupt}.}
  \bibinfo{year}{2012}\natexlab{}.
\newblock \showarticletitle{A wind power forecasting system to optimize grid
  integration}.
\newblock \bibinfo{journal}{\emph{IEEE Transactions on Sustainable Energy}}
  \bibinfo{volume}{3}, \bibinfo{number}{4} (\bibinfo{year}{2012}),
  \bibinfo{pages}{670--682}.
\newblock


\bibitem[Maldonado-Correa et~al\mbox{.}(2021)]%
        {maldonado2021wind}
\bibfield{author}{\bibinfo{person}{Jorge Maldonado-Correa}, \bibinfo{person}{JC
  Solano}, {and} \bibinfo{person}{Marco Rojas-Moncayo}.}
  \bibinfo{year}{2021}\natexlab{}.
\newblock \showarticletitle{Wind power forecasting: A systematic literature
  review}.
\newblock \bibinfo{journal}{\emph{Wind Engineering}} \bibinfo{volume}{45},
  \bibinfo{number}{2} (\bibinfo{year}{2021}), \bibinfo{pages}{413--426}.
\newblock


\bibitem[Niu et~al\mbox{.}(2022)]%
        {niu2022point}
\bibfield{author}{\bibinfo{person}{Dongxiao Niu}, \bibinfo{person}{Lijie Sun},
  \bibinfo{person}{Min Yu}, {and} \bibinfo{person}{Keke Wang}.}
  \bibinfo{year}{2022}\natexlab{}.
\newblock \showarticletitle{Point and interval forecasting of ultra-short-term
  wind power based on a data-driven method and hybrid deep learning model}.
\newblock \bibinfo{journal}{\emph{Energy}}  \bibinfo{volume}{254}
  (\bibinfo{year}{2022}), \bibinfo{pages}{124384}.
\newblock


\bibitem[Oreshkin et~al\mbox{.}(2020)]%
        {oreshkin2019n}
\bibfield{author}{\bibinfo{person}{Boris~N Oreshkin}, \bibinfo{person}{Dmitri
  Carpov}, \bibinfo{person}{Nicolas Chapados}, {and} \bibinfo{person}{Yoshua
  Bengio}.} \bibinfo{year}{2020}\natexlab{}.
\newblock \showarticletitle{N-BEATS: Neural basis expansion analysis for
  interpretable time series forecasting}.
\newblock  (\bibinfo{year}{2020}), \bibinfo{pages}{1--28}.
\newblock


\bibitem[O'Shea and Nash(2015)]%
        {o2015introduction}
\bibfield{author}{\bibinfo{person}{Keiron O'Shea} {and} \bibinfo{person}{Ryan
  Nash}.} \bibinfo{year}{2015}\natexlab{}.
\newblock \showarticletitle{An introduction to convolutional neural networks}.
\newblock \bibinfo{journal}{\emph{arXiv preprint arXiv:1511.08458}}
  (\bibinfo{year}{2015}).
\newblock


\bibitem[Perslev et~al\mbox{.}(2019)]%
        {perslev2019u}
\bibfield{author}{\bibinfo{person}{Mathias Perslev}, \bibinfo{person}{Michael
  Jensen}, \bibinfo{person}{Sune Darkner}, \bibinfo{person}{Poul~J{\o}rgen
  Jennum}, {and} \bibinfo{person}{Christian Igel}.}
  \bibinfo{year}{2019}\natexlab{}.
\newblock \showarticletitle{U-time: A fully convolutional network for time
  series segmentation applied to sleep staging}.
\newblock   \bibinfo{volume}{32} (\bibinfo{year}{2019}),
  \bibinfo{pages}{1--12}.
\newblock


\bibitem[Salinas et~al\mbox{.}(2020)]%
        {salinas2020deepar}
\bibfield{author}{\bibinfo{person}{David Salinas}, \bibinfo{person}{Valentin
  Flunkert}, \bibinfo{person}{Jan Gasthaus}, {and} \bibinfo{person}{Tim
  Januschowski}.} \bibinfo{year}{2020}\natexlab{}.
\newblock \showarticletitle{DeepAR: Probabilistic forecasting with
  autoregressive recurrent networks}.
\newblock \bibinfo{journal}{\emph{International Journal of Forecasting}}
  \bibinfo{volume}{36}, \bibinfo{number}{3} (\bibinfo{year}{2020}),
  \bibinfo{pages}{1181--1191}.
\newblock


\bibitem[Tian et~al\mbox{.}(2022)]%
        {tian2022developing}
\bibfield{author}{\bibinfo{person}{Chaonan Tian}, \bibinfo{person}{Tong Niu},
  {and} \bibinfo{person}{Wei Wei}.} \bibinfo{year}{2022}\natexlab{}.
\newblock \showarticletitle{Developing a wind power forecasting system based on
  deep learning with attention mechanism}.
\newblock \bibinfo{journal}{\emph{Energy}}  \bibinfo{volume}{257}
  (\bibinfo{year}{2022}), \bibinfo{pages}{124750}.
\newblock


\bibitem[UK~Department~for Business and Strategy(2021)]%
        {ukenergy}
\bibfield{author}{\bibinfo{person}{Energy \&~Industry UK~Department~for
  Business} {and} \bibinfo{person}{Industrial Strategy}.}
  \bibinfo{year}{2021}\natexlab{}.
\newblock \bibinfo{title}{Energy Trends}.
\newblock
\newblock
\urldef\tempurl%
\url{https://assets.publishing.service.gov.uk/government/uploads/system/uploads/attachment_data/file/1043311/Energy_Trends_December_2021.pdf}
\showURL{%
Retrieved December, 2021 from \tempurl}


\bibitem[Vaswani et~al\mbox{.}(2017)]%
        {vaswani2017attention}
\bibfield{author}{\bibinfo{person}{Ashish Vaswani}, \bibinfo{person}{Noam
  Shazeer}, \bibinfo{person}{Niki Parmar}, \bibinfo{person}{Jakob Uszkoreit},
  \bibinfo{person}{Llion Jones}, \bibinfo{person}{Aidan~N Gomez},
  \bibinfo{person}{{\L}ukasz Kaiser}, {and} \bibinfo{person}{Illia
  Polosukhin}.} \bibinfo{year}{2017}\natexlab{}.
\newblock \showarticletitle{Attention is all you need}. In
  \bibinfo{booktitle}{\emph{Advances in neural information processing
  systems}}. \bibinfo{pages}{5998--6008}.
\newblock


\bibitem[Wang et~al\mbox{.}(2011)]%
        {wang2011review}
\bibfield{author}{\bibinfo{person}{Xiaochen Wang}, \bibinfo{person}{Peng Guo},
  {and} \bibinfo{person}{Xiaobin Huang}.} \bibinfo{year}{2011}\natexlab{}.
\newblock \showarticletitle{A review of wind power forecasting models}.
\newblock \bibinfo{journal}{\emph{Energy procedia}}  \bibinfo{volume}{12}
  (\bibinfo{year}{2011}), \bibinfo{pages}{770--778}.
\newblock


\bibitem[Whittle(1963)]%
        {whittle1963prediction}
\bibfield{author}{\bibinfo{person}{Peter Whittle}.}
  \bibinfo{year}{1963}\natexlab{}.
\newblock \bibinfo{booktitle}{\emph{Prediction and regulation by linear
  least-square methods}}.
\newblock \bibinfo{publisher}{English Universities Press}.
\newblock


\bibitem[Wu et~al\mbox{.}(2022)]%
        {wu2022interpretable}
\bibfield{author}{\bibinfo{person}{Binrong Wu}, \bibinfo{person}{Lin Wang},
  {and} \bibinfo{person}{Yu-Rong Zeng}.} \bibinfo{year}{2022}\natexlab{}.
\newblock \showarticletitle{Interpretable wind speed prediction with
  multivariate time series and temporal fusion transformers}.
\newblock \bibinfo{journal}{\emph{Energy}}  \bibinfo{volume}{252}
  (\bibinfo{year}{2022}), \bibinfo{pages}{123990}.
\newblock


\bibitem[Xie et~al\mbox{.}(2018)]%
        {xie2018nonparametric}
\bibfield{author}{\bibinfo{person}{Wei Xie}, \bibinfo{person}{Pu Zhang},
  \bibinfo{person}{Rong Chen}, {and} \bibinfo{person}{Zhi Zhou}.}
  \bibinfo{year}{2018}\natexlab{}.
\newblock \showarticletitle{A nonparametric Bayesian framework for short-term
  wind power probabilistic forecast}.
\newblock \bibinfo{journal}{\emph{IEEE Transactions on Power Systems}}
  \bibinfo{volume}{34}, \bibinfo{number}{1} (\bibinfo{year}{2018}),
  \bibinfo{pages}{371--379}.
\newblock


\bibitem[Xu et~al\mbox{.}(2021)]%
        {xu2021autoformer}
\bibfield{author}{\bibinfo{person}{Jiehui Xu}, \bibinfo{person}{Jianmin Wang},
  \bibinfo{person}{Mingsheng Long}, {et~al\mbox{.}}}
  \bibinfo{year}{2021}\natexlab{}.
\newblock \showarticletitle{Autoformer: Decomposition transformers with
  auto-correlation for long-term series forecasting}.
\newblock   \bibinfo{volume}{34} (\bibinfo{year}{2021}),
  \bibinfo{pages}{0--12}.
\newblock


\bibitem[Yan et~al\mbox{.}(2015)]%
        {yan2015reviews}
\bibfield{author}{\bibinfo{person}{Jie Yan}, \bibinfo{person}{Yongqian Liu},
  \bibinfo{person}{Shuang Han}, \bibinfo{person}{Yimei Wang}, {and}
  \bibinfo{person}{Shuanglei Feng}.} \bibinfo{year}{2015}\natexlab{}.
\newblock \showarticletitle{Reviews on uncertainty analysis of wind power
  forecasting}.
\newblock \bibinfo{journal}{\emph{Renewable and Sustainable Energy Reviews}}
  \bibinfo{volume}{52} (\bibinfo{year}{2015}), \bibinfo{pages}{1322--1330}.
\newblock


\bibitem[Ye et~al\mbox{.}(2023)]%
        {ye2023web}
\bibfield{author}{\bibinfo{person}{Hangting Ye}, \bibinfo{person}{Zhining Liu},
  \bibinfo{person}{Wei Cao}, \bibinfo{person}{Amir~M Amiri},
  \bibinfo{person}{Jiang Bian}, \bibinfo{person}{Yi Chang},
  \bibinfo{person}{Jon~D Lurie}, \bibinfo{person}{Jim Weinstein}, {and}
  \bibinfo{person}{Tie-Yan Liu}.} \bibinfo{year}{2023}\natexlab{}.
\newblock \showarticletitle{Web-based long-term spine treatment outcome
  forecasting}. In \bibinfo{booktitle}{\emph{Proceedings of the 29th ACM SIGKDD
  Conference on Knowledge Discovery and Data Mining}}.
  \bibinfo{pages}{3082--3092}.
\newblock


\bibitem[Yi et~al\mbox{.}(2023a)]%
        {yi2023survey}
\bibfield{author}{\bibinfo{person}{Kun Yi}, \bibinfo{person}{Qi Zhang},
  \bibinfo{person}{Longbing Cao}, \bibinfo{person}{Shoujin Wang},
  \bibinfo{person}{Guodong Long}, \bibinfo{person}{Liang Hu},
  \bibinfo{person}{Hui He}, \bibinfo{person}{Zhendong Niu},
  \bibinfo{person}{Wei Fan}, {and} \bibinfo{person}{Hui Xiong}.}
  \bibinfo{year}{2023}\natexlab{a}.
\newblock \showarticletitle{Neural Time Series Analysis with Fourier Transform:
  A Survey}.
\newblock \bibinfo{journal}{\emph{arXiv preprint}} (\bibinfo{year}{2023}).
\newblock


\bibitem[Yi et~al\mbox{.}(2023b)]%
        {yi2023fouriergnn}
\bibfield{author}{\bibinfo{person}{Kun Yi}, \bibinfo{person}{Qi Zhang},
  \bibinfo{person}{Wei Fan}, \bibinfo{person}{Hui He}, \bibinfo{person}{Liang
  Hu}, \bibinfo{person}{Pengyang Wang}, \bibinfo{person}{Ning An},
  \bibinfo{person}{Longbing Cao}, {and} \bibinfo{person}{Zhendong Niu}.}
  \bibinfo{year}{2023}\natexlab{b}.
\newblock \showarticletitle{FourierGNN: Rethinking Multivariate Time Series
  Forecasting from a Pure Graph Perspective}. In
  \bibinfo{booktitle}{\emph{Thirty-seventh Conference on Neural Information
  Processing Systems}}.
\newblock


\bibitem[Yi et~al\mbox{.}(2023c)]%
        {yi2023frequency}
\bibfield{author}{\bibinfo{person}{Kun Yi}, \bibinfo{person}{Qi Zhang},
  \bibinfo{person}{Wei Fan}, \bibinfo{person}{Shoujin Wang},
  \bibinfo{person}{Pengyang Wang}, \bibinfo{person}{Hui He},
  \bibinfo{person}{Ning An}, \bibinfo{person}{Defu Lian},
  \bibinfo{person}{Longbing Cao}, {and} \bibinfo{person}{Zhendong Niu}.}
  \bibinfo{year}{2023}\natexlab{c}.
\newblock \showarticletitle{Frequency-domain MLPs are More Effective Learners
  in Time Series Forecasting}. In \bibinfo{booktitle}{\emph{Thirty-seventh
  Conference on Neural Information Processing Systems}}.
\newblock


\bibitem[Yildiz et~al\mbox{.}(2021)]%
        {yildiz2021improved}
\bibfield{author}{\bibinfo{person}{Ceyhun Yildiz}, \bibinfo{person}{Hakan
  Acikgoz}, \bibinfo{person}{Deniz Korkmaz}, {and} \bibinfo{person}{Umit
  Budak}.} \bibinfo{year}{2021}\natexlab{}.
\newblock \showarticletitle{An improved residual-based convolutional neural
  network for very short-term wind power forecasting}.
\newblock \bibinfo{journal}{\emph{Energy Conversion and Management}}
  \bibinfo{volume}{228} (\bibinfo{year}{2021}), \bibinfo{pages}{113731}.
\newblock


\bibitem[Yu et~al\mbox{.}(2019)]%
        {yu2019lstm}
\bibfield{author}{\bibinfo{person}{Ruiguo Yu}, \bibinfo{person}{Jie Gao},
  \bibinfo{person}{Mei Yu}, \bibinfo{person}{Wenhuan Lu},
  \bibinfo{person}{Tianyi Xu}, \bibinfo{person}{Mankun Zhao},
  \bibinfo{person}{Jie Zhang}, \bibinfo{person}{Ruixuan Zhang}, {and}
  \bibinfo{person}{Zhuo Zhang}.} \bibinfo{year}{2019}\natexlab{}.
\newblock \showarticletitle{LSTM-EFG for wind power forecasting based on
  sequential correlation features}.
\newblock \bibinfo{journal}{\emph{Future Generation Computer Systems}}
  \bibinfo{volume}{93} (\bibinfo{year}{2019}), \bibinfo{pages}{33--42}.
\newblock


\bibitem[Yuan et~al\mbox{.}(2017)]%
        {yuan2017wind}
\bibfield{author}{\bibinfo{person}{Xiaohui Yuan}, \bibinfo{person}{Qingxiong
  Tan}, \bibinfo{person}{Xiaohui Lei}, \bibinfo{person}{Yanbin Yuan}, {and}
  \bibinfo{person}{Xiaotao Wu}.} \bibinfo{year}{2017}\natexlab{}.
\newblock \showarticletitle{Wind power prediction using hybrid autoregressive
  fractionally integrated moving average and least square support vector
  machine}.
\newblock \bibinfo{journal}{\emph{Energy}}  \bibinfo{volume}{129}
  (\bibinfo{year}{2017}), \bibinfo{pages}{122--137}.
\newblock


\bibitem[Zhou et~al\mbox{.}(2021)]%
        {zhou2021informer}
\bibfield{author}{\bibinfo{person}{Haoyi Zhou}, \bibinfo{person}{Shanghang
  Zhang}, \bibinfo{person}{Jieqi Peng}, \bibinfo{person}{Shuai Zhang},
  \bibinfo{person}{Jianxin Li}, \bibinfo{person}{Hui Xiong}, {and}
  \bibinfo{person}{Wancai Zhang}.} \bibinfo{year}{2021}\natexlab{}.
\newblock \showarticletitle{Informer: Beyond efficient transformer for long
  sequence time-series forecasting}. In \bibinfo{booktitle}{\emph{Proceedings
  of AAAI}}.
\newblock


\end{thebibliography}

% \appendix

\end{document}